\documentclass[a4paper,11pt,preprintnumbers]{article}

\pdfoutput=1 

\usepackage{jcappub} 

\usepackage[T1]{fontenc} 

\usepackage[table, dvipsnames]{xcolor}

\newcommand{\cH}{\mathcal{H}}

\newcommand{\eV}{ \ {\rm eV} }
\newcommand{\keV}{ \ {\rm keV} }

\title{Enhanced Matter Power Spectrum\\from Axion Kination after Big Bang Nucleosynthesis}

\author[1]{Raymond T.~Co,}
\author[2]{Nicolas Fernandez,}
\author[3]{Akshay Ghalsasi}
\author[4,5,6]{Keisuke~Harigaya}
\author[7]{Jessie Shelton}

\affiliation[1]{Physics Department, Indiana University, Bloomington, IN 47405, USA}
\affiliation[2]{NHETC, Department of Physics and Astronomy, Rutgers University, Piscataway, NJ 08854, USA}
\affiliation[3]{Jefferson Physical Laboratory, Harvard University, Cambridge, MA 02138, USA}
\affiliation[4]{Department of Physics, University of Chicago, Chicago, IL 60637, USA}
\affiliation[5]{Enrico Fermi Institute, Leinweber Institute for Theoretical Physics, and Kavli Institute for Cosmological Physics, University of Chicago, Chicago, IL 60637, USA}
\affiliation[6]{Kavli Institute for the Physics and Mathematics of the Universe (WPI),
The University of Tokyo Institutes for Advanced Study,
The University of Tokyo, Kashiwa, Chiba 277-8583, Japan}
\affiliation[7]{Illinois Center for Advanced Studies of the Universe and Department of Physics, University of Illinois at Urbana-Champaign, Urbana, IL 61801, USA}

\emailAdd{rco@iu.edu}
\emailAdd{nico.fer@rutgers.edu}
\emailAdd{aghalsasi@fas.harvard.edu}
\emailAdd{kharigaya@uchicago.edu}
\emailAdd{sheltonj@illinois.edu}

\abstract{Despite stringent constraints from Big Bang Nucleosynthesis (BBN) and cosmic microwave background (CMB) observations, it is still possible for well-motivated particle physics models to substantially alter the cosmic expansion history between BBN and recombination. In this work we consider two different axion models that can realize a period of first matter domination, then kination, in this epoch. We perform fits to both primordial element abundances as well as CMB data and determine that up to a decade of late axion domination is allowed by these probes of the early universe.  We establish the implications of late axion domination for the matter power spectrum on the scales $1/\mathrm{Mpc}\lesssim k \lesssim 10^3/$Mpc. Our `log' model predicts a relatively modest bump-like feature together with a small suppression relative to the standard $\Lambda$CDM predictions on either side of the enhancement. Our `two-field' model predicts a larger, plateau-like feature that realizes enhancements to the matter power spectrum of up to two orders of magnitude.  These features have interesting implications for structure formation at the forefront of current detection capabilities.
}

\begin{document}

\maketitle

\section{Introduction}
\label{sec:intro}

The thermal history of our early universe is anchored by two observational probes: Big Bang Nucleosynthesis (BBN) and the Cosmic Microwave Background (CMB). 
The overall  agreement of these two suites of measurements with predictions of the standard $\Lambda$CDM cosmological model is excellent, and together they provide a detailed test of the thermal history of our universe up to temperatures of order 5 MeV. 
Critically, the concordance between BBN and CMB determinations of both the baryon-to-photon ratio $\eta_B$ and the  number of effective relativistic degrees of freedom $N_\mathrm{eff}$ places very tight constraints on the possible evolution of the universe between these two epochs \cite{Yeh:2022heq,Sobotka:2022vrr, Sobotka:2023bzr}.   In particular, since observations do not allow for a substantial injection of entropy after the beginning of BBN, it is impossible to accommodate an abundance of a metastable species that is large enough to contribute substantially to the Hubble rate at these times.  The avenues for sizeable departures from the standard expansion history at late, i.e., post-BBN times, are thus extremely limited.

Coherent scalar fields with a time-dependent equation of state  are one major exception  \cite{Co:2021lkc,Sobotka:2024tat}. Given a suitable equation of state $w(a)$, such fields can realize a  transition from redshifting more slowly than radiation, such that the scalar field can come to dominate, to redshifting faster than radiation, such that its contribution to the Hubble rate rapidly becomes negligible. One particularly interesting class of such models is a complex field containing an axion that rotates, rather than oscillates, in a wine-bottle potential \cite{Co:2019wyp}.  In these theories, a $U(1)$ charge density is carried by a complex field circularly rotating in field space. The conservation of $U(1)$ charge implies that the radius of the rotation decreases with Hubble as the universe expands. When the field is rotating while away from the minimum of its potential, its energy density redshifts like matter, $\rho \propto a^{-3}$, where we assume a nearly quadratic potential of the radial direction of the rotating field.  Once the axion reaches its minimum, however, it behaves as kination, $\rho \propto a^{-6}$.  This axion kination cosmology naturally provides baryogenesis mechanisms \cite{Co:2019wyp,Domcke:2020kcp,Co:2020xlh,Co:2020jtv,Harigaya:2021txz,Chakraborty:2021fkp,Kawamura:2021xpu,Co:2021qgl,Co:2022aav,Barnes:2022ren,Co:2022kul,Berbig:2023uzs,Chun:2023eqc,Barnes:2024jap,Wada:2024cbe,Datta:2024xhg}, opens up new parameter space for axion dark matter \cite{Co:2019jts,Co:2020dya,Eroncel:2022vjg,Eroncel:2025qlk,Bodas:2025eca}, and can allow for gravitational-wave windows onto the evolution of our universe at very early times \cite{Co:2021rhi,Madge:2021abk,Gouttenoire:2021wzu,Harigaya:2023mhl,Bodas:2025wef}. Critically, axion kination also allows for an early matter-dominated era {\em after} BBN: the matter-like behavior of the axion at early times allows it to come to dominate over the Standard Model radiation bath, while its natural transition to kination-like behavior ensures that this axion domination ends without entropy injection. Axion kination models can drive baryogenesis even when the matter-to-kination transition occurs post-BBN~\cite{Co:2024oek}.

A period of early matter domination post-BBN has a sizeable impact on the growth of structure on scales that experience the modified expansion history.  The scales entering the horizon between BBN and CMB epochs, $0.5 /\mathrm{Mpc} \lesssim k\lesssim 100/ $Mpc, straddle the boundary between dark and luminous halos, and thus are some of the scales of the most interest for understanding small-scale structure and the early growth of galaxies.  
Current and upcoming surveys are poised to have a transformational impact on our understanding of the matter power spectrum (MPS) on these scales, particularly through the wide range of probes enabled by observations in the optical instruments Roman, Euclid, and Rubin  and the infrared observations from JWST \cite{Bechtol:2022koa, Mao:2022fyx}.  Indeed  early results from JWST may already contain intriguing hints toward enhanced early structure formation, such as the abundance of high-redshift galaxies \cite{2022ApJ...940L..14N,2022ApJ...938L..15C,2024Natur.633..318C,2023MNRAS.519.1201A,2023Natur.616..266L} as well as the unexpectedly efficient formation of supermassive black holes \cite{2025arXiv250821748J,2024NatAs...8..126B, 2023ApJ...957L...7K,2023ApJ...953L..29L,2024Natur.628...57F,2023MNRAS.525.1353J,2024arXiv241018763N,Qin:2025ymc}.

In this work we consider two different models of axion kination, which make slightly different predictions for the axion equation of state $w(a)$, and establish the consequences of  axion domination after BBN. The different microphysical behavior of the two models gives rise to substantially different predictions for the small-scale MPS: one produces a relatively localized bump, and one predicts a larger plateau-like feature.   In both cases a notable enhancement to the linear MPS can be obtained on scales relevant for dwarf galaxies and below.  

The organization of this paper is as follows: in Sec.~\ref{sec:cosmology} we briefly review the cosmology of axion kination and introduce the two specific models we consider in this work. In Sec.~\ref{sec:pertgrowth} we discuss the evolution of cosmological perturbations during late axion  domination, and highlight how the different axion equations of state in our two models give rise to different behavior. Sec.~\ref{sec:constraints} establishes BBN and CMB constraints on the duration of axion domination, while Sec.~\ref{sec:SSMPS} shows fiducial power spectra for both models together with a discussion of observables. We conclude in Sec.~\ref{sec:summary}. In Appendix~A we provide additional detail about the parametric dependence of the axion-induced features in the MPS.

\section{Axion Kination Cosmology}
\label{sec:cosmology}

The phenomenology we consider in this work relies on the existence of matter-dominated ($\rho \propto a^{-3}$) and kination-dominated ($\rho \propto a^{-6}$) epochs that occur after Big Bang Nucleosynthesis (BBN) and before recombination. Since the kination energy density redshifts faster than that of radiation ($\rho \propto a^{-4}$), no entropy production is needed to return to radiation domination from matter domination, which allows such eras to exist after BBN. As we will review in this section, such a scenario is well motivated by the early-universe dynamics of a complex field with a $U(1)$ symmetry, e.g., those containing axion-like particles in the angular degree of freedom. Specifically, a complex field rotating in field space gives rise to this unique matter-like, then kination-like evolution of its energy density.

\begin{figure}
\centering   \includegraphics[width=0.6\linewidth]{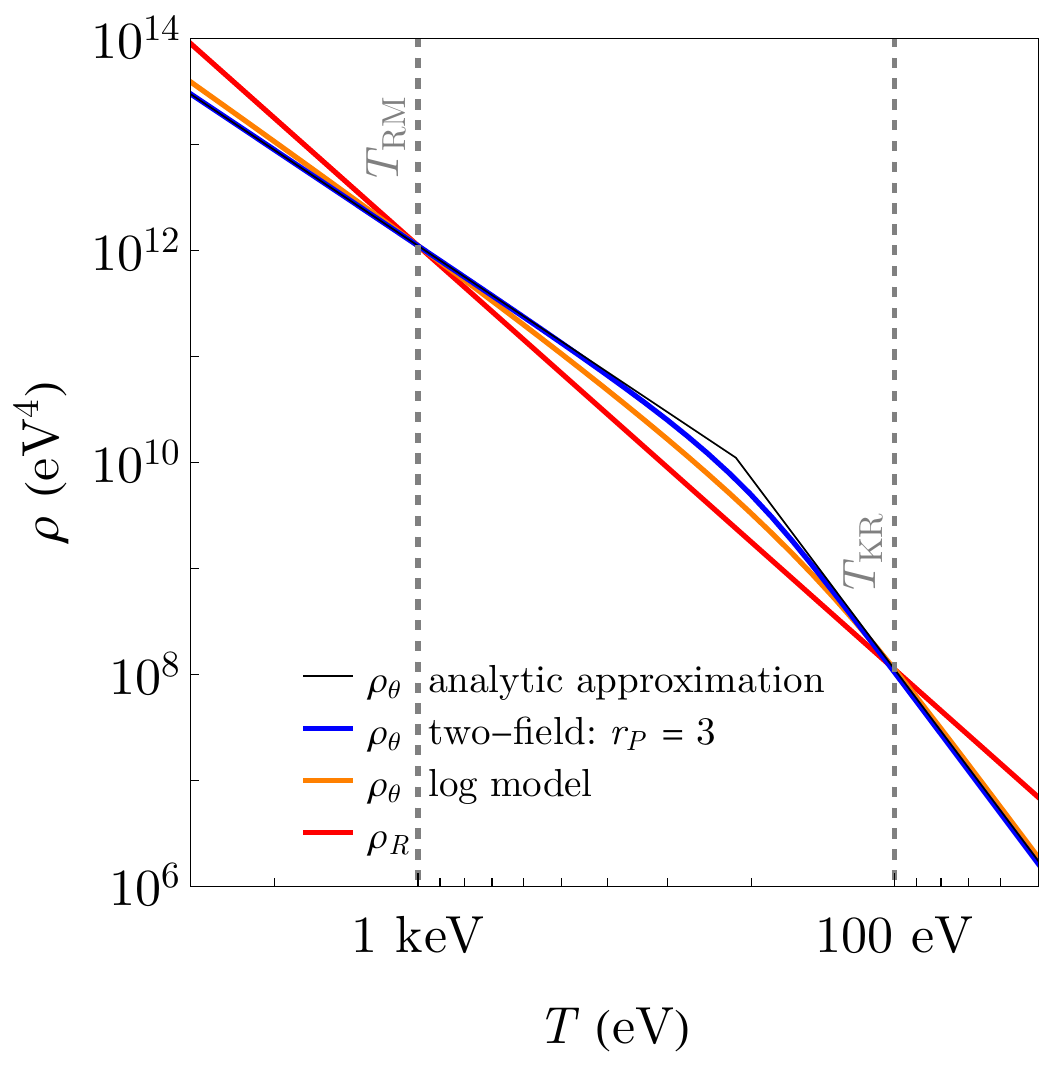}
\caption{Energy density as a function of (decreasing) temperature $T$, for radiation (red), the two-field model with $r_P=3$ (blue), and the log model (orange). The black line shows  scalings of $\rho_\theta \propto T^3$ and $\rho_\theta \propto T^6$ at  early and late times, respectively.}
\label{fig:rho}
\end{figure}

Complex-field rotations were first introduced in the context of Affleck-Dine baryogenesis~\cite{Affleck:1984fy,Dine:1995kz}, where the field rotates in a potential that is nearly quadratic in the radial direction and the energy density simply scales as matter. An interesting scenario arises when the potential instead features spontaneous $U(1)$-symmetry breaking. A complex field rotation at the minimum of a wine-bottle potential leads to an energy density that scales like kination. If the potential is nearly quadratic in the limit when the field value is much larger than the vacuum expectation value (vev), the energy density scales as matter. This implies that a field rotation starting with large initial radius will evolve through the matter and kination scaling~\cite{Co:2019wyp}. The initiation of field rotations in the Affleck-Dine mechanism in fact requires a large initial radius, where  higher-dimensional explicit $U(1)$-breaking operators can generate the ``kick'' in the angular direction when the complex field starts to evolve in the potential.

The rotation initiated by the Affleck-Dine mechanism is necessarily elliptic because both radial and angular modes receive kicks. Elliptic motion does not exhibit matter-kination scaling, since it contains radial oscillations that continue to behave as matter even after the radius of the rotation becomes as small as the vev. However, the motion becomes circular once the complex field is thermalized via scattering with particles in the thermal radiation bath; in thermal equilibrium, the axion field takes the form that has the minimal energy at fixed $U(1)$ charge, namely, circular rotation~\cite{Co:2019wyp,Domcke:2022wpb}.%
\footnote{
If the axion continues to couple to the thermal bath even after the QCD phase transition, axion dark radiation would be overproduced. This overproduction is avoided when thermalization occurs via scattering with a particle heavier than the QCD scale, in which case the axion will decouple from the thermal bath before the QCD phase transition. 
See~\cite{Co:2024oek} for a concrete thermalization model for a low-scale rotating axion that also achieves baryogenesis (axiogenesis).    
}

The scaling of the energy densities of the radiation and the rotation is sketched in Fig.~\ref{fig:rho}. Here we show a simple joined power-law description of matter, then kination domination in black, together with the specific predictions of the two models we detail below. 
We denote the early matter-radiation equality as ``RM'' and the kination-radiation equality as ``KR''.
We now review two models where the potential of the $U(1)$ symmetry-breaking field is nearly quadratic at large field values and thus enables this axion kination cosmology.

\subsection{Log Potential Model}

We denote the $U(1)$ symmetry-breaking field as $P$ and decompose it into a radial degree of freedom $S$ and an angular degree of freedom $\theta$,
\begin{equation}
    P = \frac{1}{\sqrt{2}}S e^{i \theta}.
\end{equation}
In supersymmetric theories, the potential of $P$ may be suppressed and arise only via supersymmetry breaking. At  tree-level, the potential is quadratic, but quantum corrections from the interaction of $P$ with other fields can make the quadratic term logarithmically depend on the field value of $P$. In particular, starting from a positive mass-squared at UV scales, the mass-squared may become negative at IR scales, for which the vev of $P$ is nonzero~\cite{Moxhay:1984am}. The decay constant in this case is obtained by dimensional transmutation. We parametrize the potential as
\begin{align}
    V(P) = \frac{1}{2} m^2 |P|^2 \left({\rm ln}\frac{2|P|^2}{f_a^2} -1\right),
\end{align}
where the vev of $S$ is $f_a$.

The evolution of the circular motion can be derived in the following way~\cite{Co:2019wyp}. Neglecting the Hubble expansion rate  $H$ and  ${\rm d}S/{\rm d}t$, the equation of motion of $S$ gives $({\rm d\theta}/{{\rm d}t})^2 = (\partial V/\partial S)/S$. For this approximation to be consistent, we must have $|{\rm d\theta}/{{\rm d}t}| \gg H$.
The equation of state parameter $w = p/\rho$ depends on the radius of the rotation and is given by
\begin{align}
 w & = -1 + \frac{2\ln(r^2)}{2 \ln(r^2) -1 + r^{-2}},~~
 r^2  \equiv \left(\frac{S}{f_a}\right)^2 .
\end{align}
Note that $w\sim 0$ for $r \gg 1$ and $w \simeq 1$ for $r \simeq 1$, achieving the matter-kination scaling.

\begin{figure}
\centering   \includegraphics[width=0.7\linewidth]{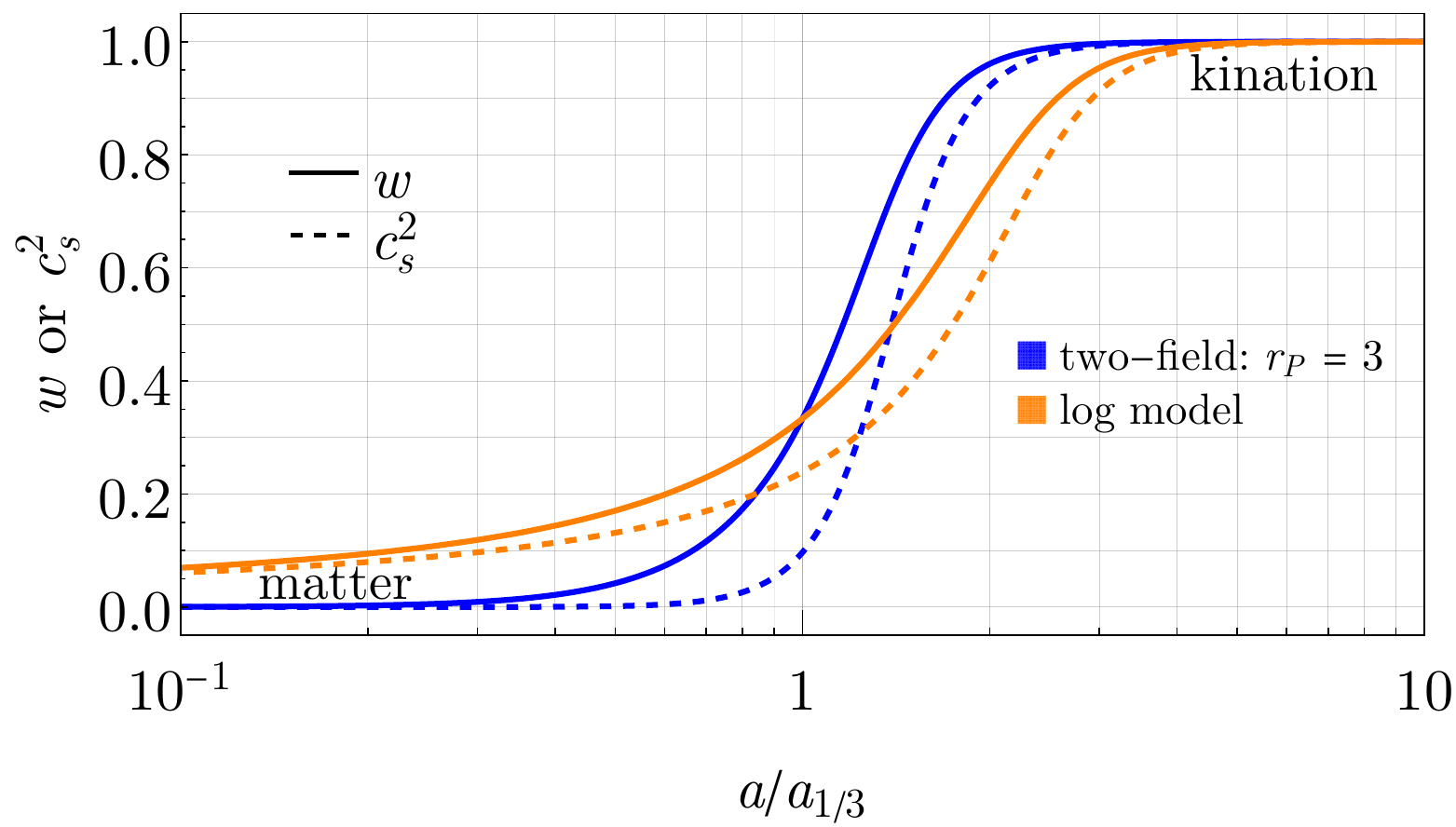}
\caption{Equation of state $w$ (solid line) and speed of sound squared $c_s^2$ (dashed line) as a function of scale factor $a$ normalized to the scale factor $a_{1/3}$ when $w=1/3$, for the two-field model with $r_P=3$ (blue) and the log model (orange). The equation of state corresponding to the black line in Fig.~\ref{fig:rho}  would be a unit step function centered at $a/a_{1/3}=1$.
}
\label{fig:wcs2log}
\end{figure}

The evolution of $w$ is determined by that of $r$, whose change is negligible over each rotation cycle but evolves adiabatically due to cosmic expansion.
The evolution of $r$ as a function of the scale factor $a$ can be derived by the equation of motion of $\theta$, which is nothing but charge conservation $({\rm d\theta}/{{\rm d}t}) S^2 \propto a^{-3}$. We find
\begin{align}
 r^4  = \frac{ 4 \left(\frac{a_{1/3}}{a}\right)^6\frac{4+2 W(-\frac{2}{e^2})}{W^2(-\frac{2}{e^2})}}{ W\left( 4\left(\frac{a_{1/3}}{a}\right)^6 \frac{4+2 W(-\frac{2}{e^2})}{W^2(-\frac{2}{e^2})}\right)},
\end{align}
where $W$ is the Lambert function and $a_{1/3}$ is the scale factor at which $w = 1/3$. For a given initial field value $S_i$ or initial temperature $T_i$ at $a_i \ll a_{1/3}$,  $a_{1/3}$ is expressed as 
\begin{align}
    a_{1/3} & = a_i \left( \frac{S_i}{f_a} \right)^{2/3} \left[ \ln\left(S_i/f_a \right) \right]^{1/6} \left( \frac{W^2(-\frac{2}{e^2})}{4+2 W(-\frac{2}{e^2})} \right)^{\frac{1}{6}} \\
    & \simeq 0.61~ a_i \left( \frac{S_i}{f_a} \right)^{2/3} \left[ \ln\left(S_i/f_a \right) \right]^{1/6} . \nonumber 
\end{align}
The resulting sound speed can be expressed as $c_s^2 = w-({\rm d}w/{\rm d}r) ({\rm d}r/{\rm d ln}a) /(3(1+w))$.
In Fig.~\ref{fig:wcs2log}, the orange curves show the evolution of $w$ and $c_s^2$ for the log-potential model as a function of $a$. Both $w$ and $c_s^2$ change gradually from $0$ to $1$.

\begin{figure}
\centering   \includegraphics[width=0.495\linewidth]{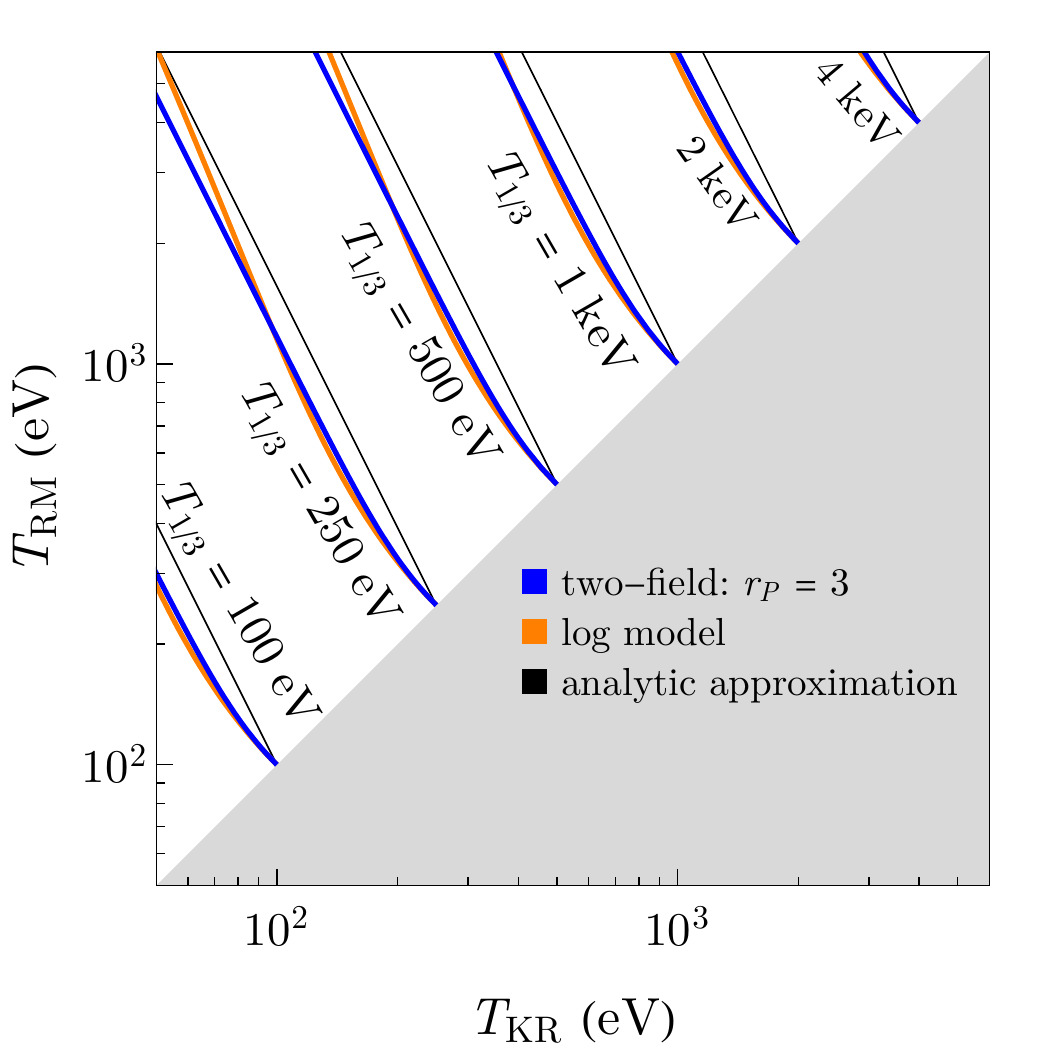}
\includegraphics[width=0.495\linewidth]{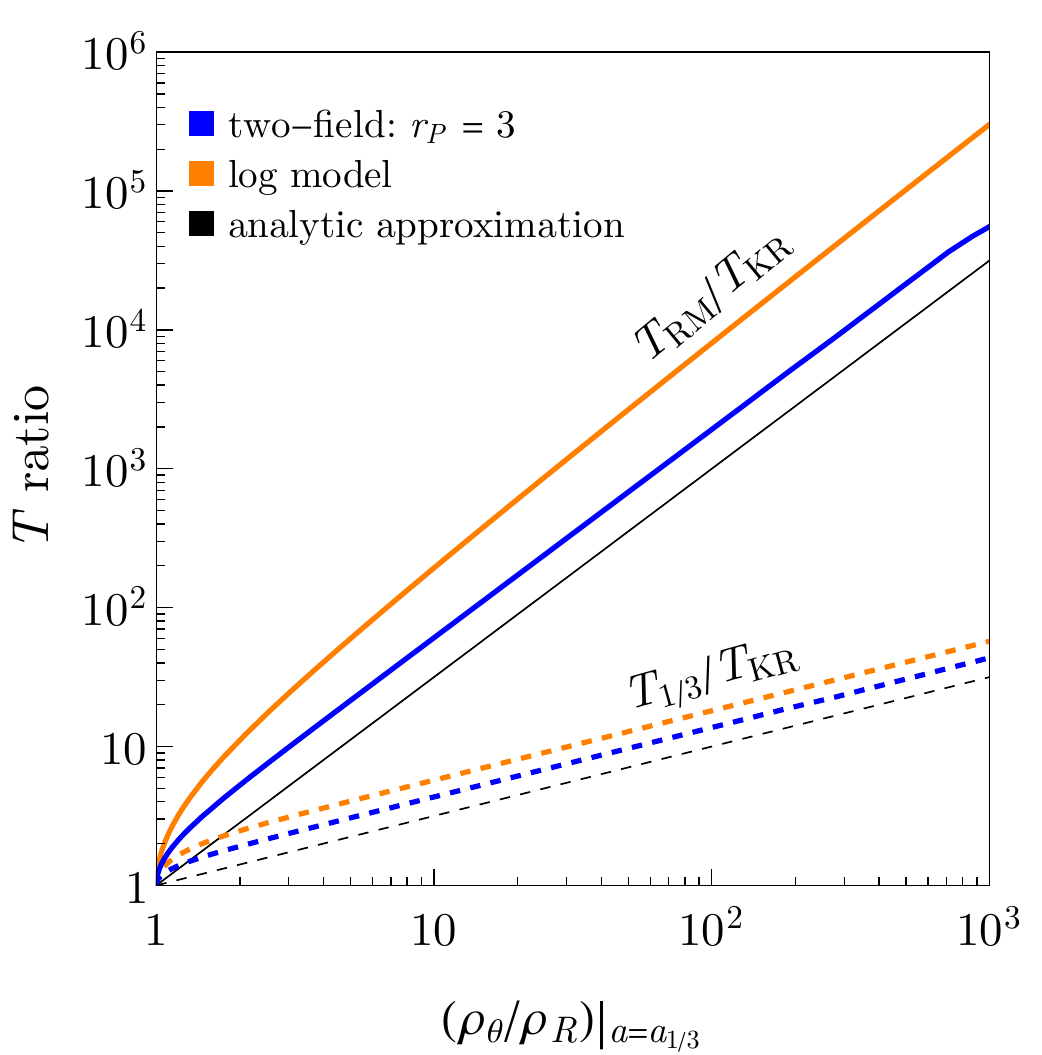}
\caption{In both panels, blue (orange) curves are for two-field (log) model, while the black lines show the approximate scalings~\cite{Co:2021lkc}. \textbf{Left}: The temperature $T_{1/3}$ at which $w=1/3$ as a function of the radiation-to-matter domination transition temperature $T_{\rm RM}$ and the kination-to-radiation domination transition $T_{\rm KR}$ temperature. \textbf{Right}: Temperature ratios as functions of the ratio of the rotation energy density to that of radiation when $w=1/3$. The black lines show the scaling of the $T$ ratio $\propto (\rho_\theta / \rho_R |_{a = a_{1/3}})^{3/2}$ (solid) and $(\rho_\theta / \rho_R |_{a = a_{1/3}})^{1/2}$ (dashed).
}
\label{fig:T_log}
\end{figure}

The left panel of Fig.~\ref{fig:T_log} shows the contours of $T(a_{1/3}) \equiv T_{1/3}$. For $T_{\rm RM} \gg T_{\rm KR}$, $T_{1/3}$ approximately scales as $T_{\rm RM}^{1/2} T_{\rm KR}^{2/3}$, but as  $T_{\rm RM}$ approaches $T_{\rm KR}$, the scaling no longer holds. The right panel shows the relation between the ratios of the temperatures and the maximum value of $\rho_\theta/\rho_R$ with $\rho_R$ the radiation energy density, which occurs at $a= a_{1/3}$.

\subsection{Two-Field Model}

Another type of supersymmetric model involves two $U(1)$-charged fields $P$ and $\bar{P}$ with opposite $U(1)$ charges. Their evolution is determined by the superpotential and soft supersymmetry breaking masses,
\begin{align}
    W = \lambda X (P \bar{P} - v^2),~~ V_{\rm soft} = m_P^2 |P|^2 + m_{\bar{P}}^2 |\bar{P}|^2,
\end{align}
where $X$ is a chiral field. The $F$ term of $X$ fixes $P$ and $\bar{P}$ on a moduli space $P \bar{P} = v^2$, along which the potential is flat in the supersymmetric limit. The flat direction is lifted by the soft-supersymmetry breaking potential $V_{\rm soft}$ to give a nearly quadratic potential. 

To analyze the dynamics of the model, it is most convenient to integrate out one of $P$ and $\bar{P}$ using $P \bar{P} = v^2$ to obtain the effective Lagrangian
\begin{equation}
\label{eq:Leff}
    {\cal L}_{\rm eff} = - \left( 1 + \frac{v^4}{|P|^4}\right) \partial_\mu P^\dag \partial^\mu P -  \left( 1 + r_P^2 \frac{v^4}{|P|^4} \right) m_P^2|P|^2,
\end{equation}
where $r_P \equiv m_{\bar{P}} / m_P$.
The equation of motion of $P$ can be solved in a similar way as the log potential model while taking into account the non-canonical kinetic term of $P$. Details can be found in~\cite{Harigaya:2023mhl,Co:2024oek}.

In Fig.~\ref{fig:wcs2log}, the blue curves show the evolution of $w$ and $c_s^2$ for the two-field model as a function of $a$. One can see that $w$ and $c_s^2$ stay close to zero until $a$ approaches $a_{1/3}$, around which they change quickly to $1$.
The change becomes more rapid as $r_P$ approaches unity \cite{Co:2021lkc}. For $r_P=1$, the potential has an exchange symmetry between $P$ and $\bar{P}$, which makes $r_P \sim 1$ a natural choice of the parameter, rather than a fine-tuned choice. 
In this work for definiteness we adopt the reference value $r_P = 3$ throughout. Values of $r_P>3$ have almost identical evolution of $w$ and $c_s$ as $r_P=3$ \cite{Co:2021lkc, Bodas:2025eca} and thus predict essentially the same enhancement as $r_P=3$. 
Taking $r_P$ closer to $1$ will give rise to larger imprints on the matter power spectrum by up a factor of few in comparison with $r_P=3$, as we discuss in more detail in Sec.~\ref{sec:pertgrowth}. 

While the axion equation of state $w(a)$ realized in these two models is qualitatively similar, the differences in the evolution of $w(a)$ and thus $c^{2}_{s} (a)$  during the transition from matter to kination have important implications for the growth of perturbations during axion domination.
In particular, in the log model both $w(a)$ and $c^{2}_{s} (a)$ are appreciably nonzero during the ``matter-dominated'' phase of evolution $a\ll a_{1/3}$, while in the two-field model, both $w(a)$ and $c^{2}_{s} (a)$ are much closer to zero at early times. As we will see in the next section, this difference in the equation of state will yield strikingly different predictions for the growth of DM perturbations.

\section{Evolution of Cosmological Perturbations }
\label{sec:pertgrowth}

The complex scalar field $P$ contains two real scalar fields and its perturbations  generically have two modes. After the rotation becomes circular due to thermalization, the two modes are decomposed into a gapped (massive) mode, which is mainly composed of the radial excitation, and a gapless (massless) mode, which is nothing but the Nambu-Goldstone boson associated with the spontaneous breaking of the $U(1)$ symmetry and the time-translational symmetry into a diagonal subgroup.  The gapped mode may be integrated out to obtain the Boltzmann equations governing the evolution of scalar perturbations in the rotating axion field, which can be expressed as those of one adiabatic fluid with a time-varying equation of state~\cite{Co:2021lkc,Harigaya:2023mhl,Co:2024oek}.  
We denote the fractional density perturbation in the rotating axion as $\delta_\theta$ and the corresponding velocity perturbation as $\Theta_\theta$. In conformal Newtonian gauge, axion perturbations are described by%
\footnote{
The sound speed that appears in the Boltzmann equations is the adiabatic sound speed, coinciding with the sound speed obtained in Sec.~\ref{sec:cosmology}. This is because the axion rotation is an adiabatic fluid; once the local charge density is specified, the energy density and the pressure are uniquely determined.
}
\begin{align}
\label{eq:fluid1}
\dot \delta_\theta &= -(1+w)(\Theta_\theta -3\dot \phi)-3\cH(c_s^2 - w) \delta_\theta, \\
\label{eq:fluid2}
\dot\Theta_\theta &=  - \cH (1-3w)\Theta_\theta - \frac{\dot w}{1+w}\Theta_\theta + \frac{c_s^2}{1+w} k^2 \delta_\theta + k^2 \phi ,
\end{align}
where $k$ is the comoving wavenumber of the perturbation, dots indicate derivatives with respect to conformal time, $\cH$ is the conformal Hubble parameter, and $\phi$ is the metric perturbation. (We follow the conventions in \cite{Ma:1995ey} for metric perturbations and  neglect anisotropic stress). At the scales of interest to us, the axion interacts only gravitationally with the other components of the universe.

To establish the consequences of rotating-axion domination between BBN and CMB epochs,
we implement the axion equation of state $w(a)$, sound speed $c_s(a)$, and the axion perturbation equations in Eqs.~\eqref{eq:fluid1} and~\eqref{eq:fluid2} in CLASS~\cite{Blas:2011rf}. We numerically solve the resulting suite of coupled Boltzmann equations, assuming all perturbations are adiabatic and setting initial conditions deep in radiation domination.

While both models discussed in Sec.~\ref{sec:cosmology} give rise to enhanced power on small scales, the different equations of state $w(a)$ in the two models yield substantially different predictions for both the magnitude and shape of the enhancement. In this section we establish the general properties of the small-scale features in the MPS that are predicted by each model, and demonstrate how the size and location of these features depend on the beginning ($T_{\rm RM}$) and the ending ($T_{\rm KR}$) of axion domination. 
In order to more clearly illustrate the features that late-time axion domination imprints on the small scale MPS, in this section we will occasionally discuss values of  $T_{\rm RM}$ and $T_{\rm KR}$ that are incompatible with observations of either primordial element abundances or CMB power spectra. We discuss the observational constraints on $T_{\rm RM}$ and $T_{\rm KR}$ in Sec.~\ref{sec:constraints}.

\subsection{Two-Field Model}

\begin{figure}
\centering   
    \includegraphics[width=0.495\linewidth]{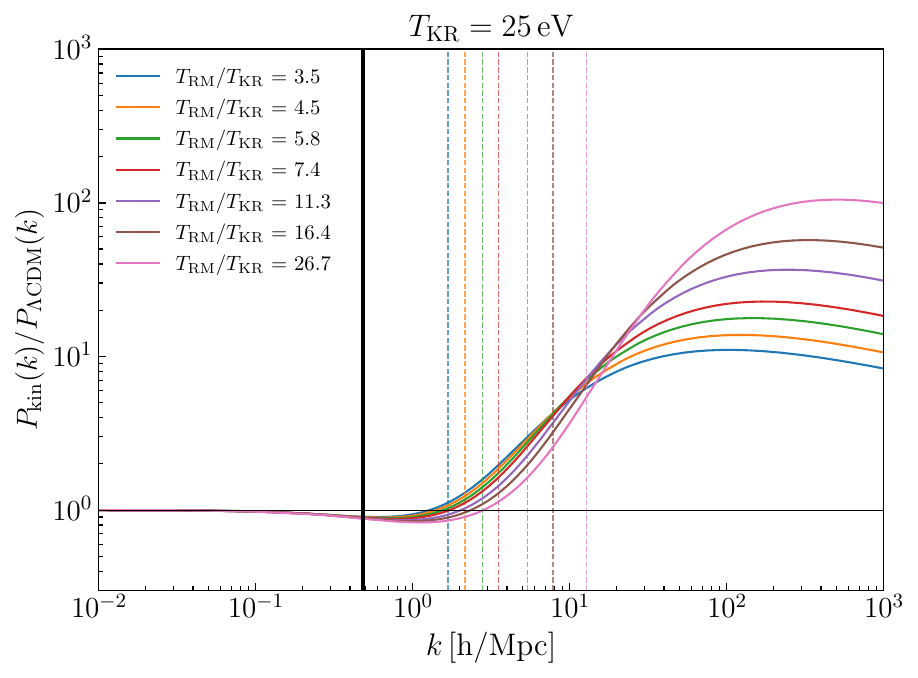}
    \includegraphics[width=0.495\linewidth]{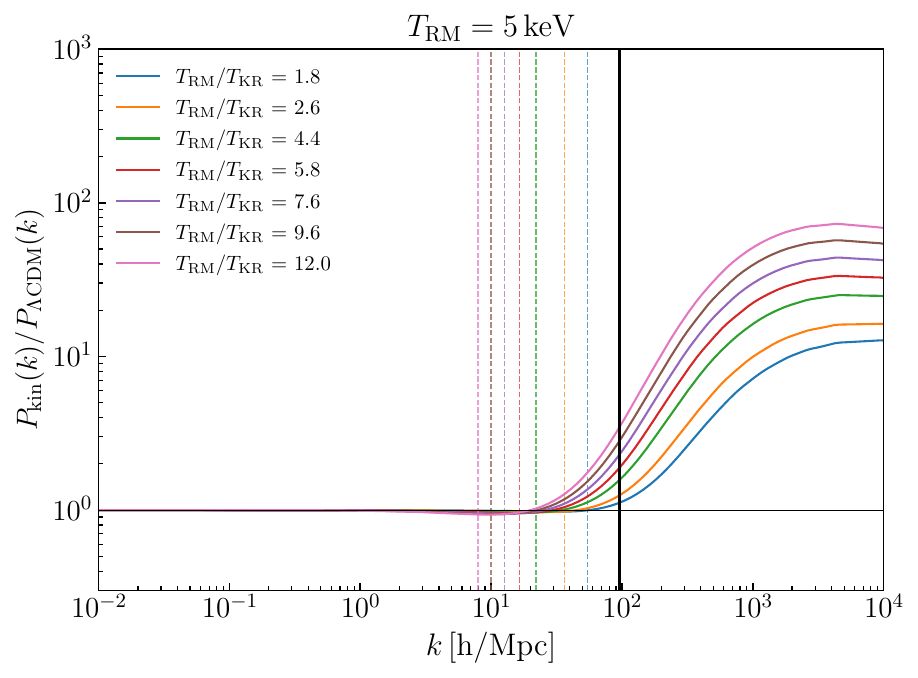}
\caption{\textbf{Left}: Ratio of the matter power spectrum  in the two-field model with $r_{\rm P} = 3$ to the prediction of the standard cosmology ($T(k)  = P_{\rm kin}(k)/P_{\rm \Lambda CDM}(k)$) with fixed $T_{\rm KR} = 25 \eV$ and varying $T_{\rm RM}$; colors correspond to different total durations of axion domination as indicated in the legend. The black vertical line indicates $k_{\rm KR}$, while colored dotted vertical lines indicate $k_{\rm RM}$ for the correspondingly colored curve.  \textbf{Right}: Same as left except for  fixed $T_{\rm RM} = 5 \keV$ with varying $T_{\rm KR}$. Here the black line indicates $k_{\rm RM}$, while  colored dotted lines indicate $k_{\rm KR}$.}
\label{fig:2fieldMPS}
\end{figure}

In this model the axion equation of state $w(a)$ makes a relatively quick transition from $w=0$ at $a\ll a_{1/3}$ to $w=1$ at $a\gg a_{1/3}$, and its derivative $\dot w(a)$ is non-negligible over a relatively narrow range of scale factors.  
In Fig.~\ref{fig:2fieldMPS} we show the resulting imprint of axion domination on the MPS. 
The left panel shows the shape of the feature for a fixed endpoint of axion domination, $T_{\rm KR} = 25$ eV, with varying starting point $T_{\rm RM}$, while the right panel shows results for a fixed $T_{\rm RM} =5$ keV and varying endpoint $T_{\rm KR}$. For the remaining cosmological parameters we adopt the Planck 2018 best-fit values as given in CLASS Planck 2018 best-fit input file \cite{Blas:2011rf}. 

As Fig.~\ref{fig:2fieldMPS} shows, the two-field model predicts a broad enhancement to the MPS for $k\gtrsim k_{\rm KR}$ 
that falls off only gradually at high $k$. The size of the enhancement can be well over an order of magnitude, as modes that are within the horizon for the entire duration of axion domination experience two phases of growth, as we now discuss.
  
\begin{figure}
\centering   \includegraphics[width=0.495\linewidth]{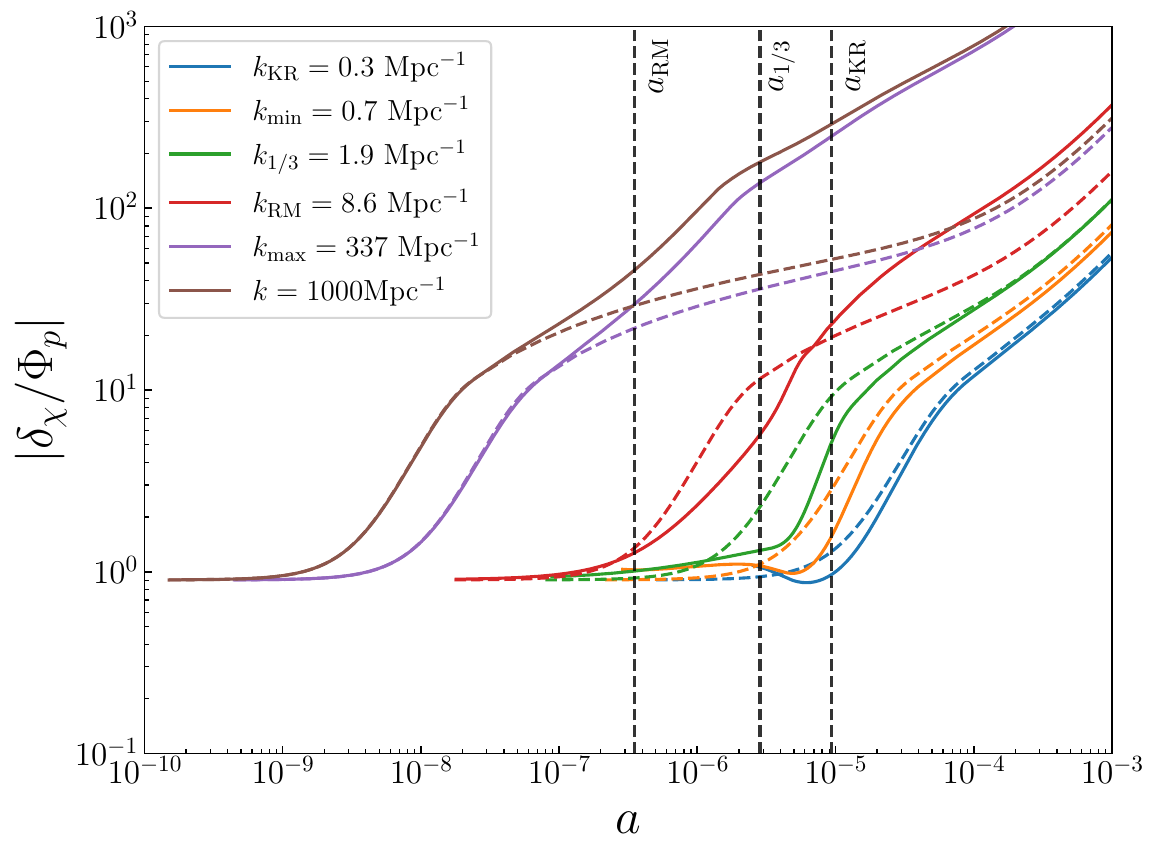}
\includegraphics[width=0.495\linewidth]{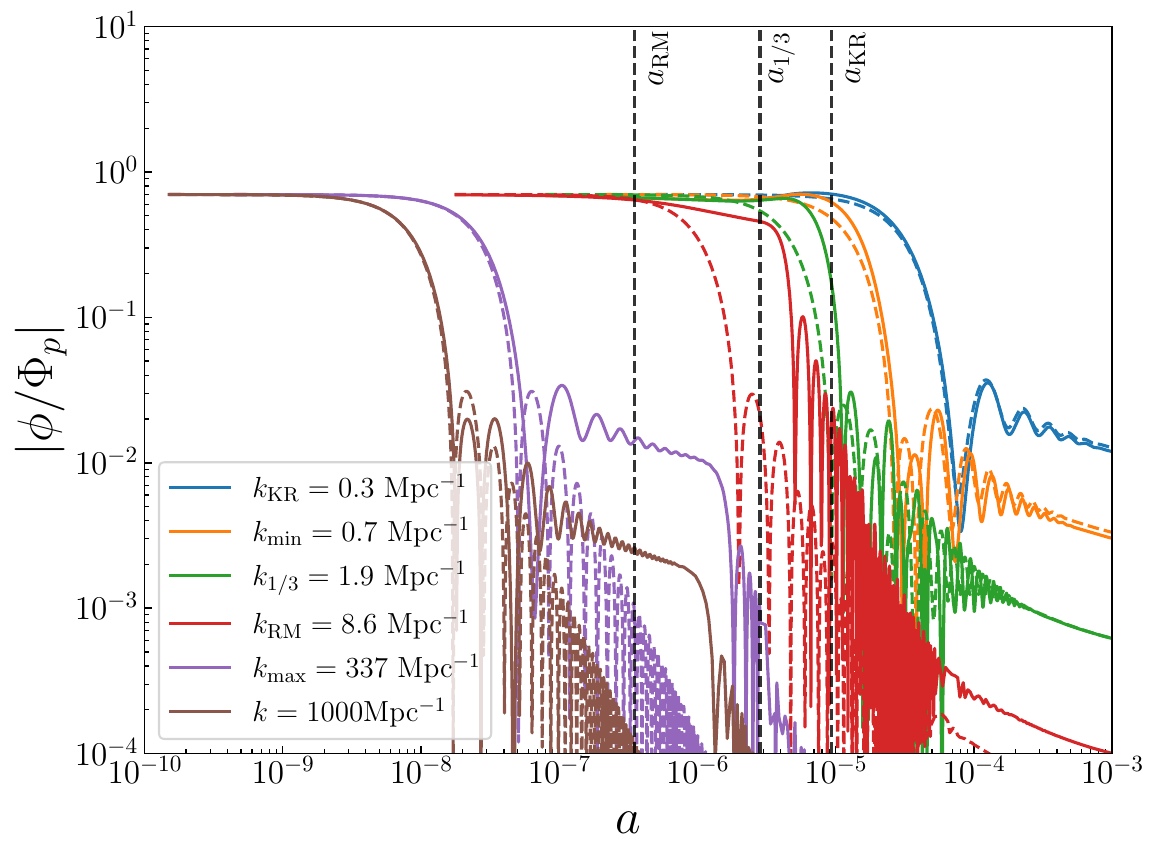}
\caption{Evolution of conformal Newtonian gauge CDM perturbations $\delta_\chi$ (left) and metric perturbations $\phi$ (right) normalized to $\Phi_p$, the primordial curvature perturbation, with scale factor in a cosmology with axion domination in the two-field model (solid) and $\Lambda \rm{CDM}$ (dashed), for  $T_{\rm KR} = 25 \eV $ and $T_{\rm RM} = 26.7~ T_{\rm KR} = 0.67$ keV. Colors indicate different comoving wavenumbers $k$ as shown in the legend. }

\label{fig:twofieldpert}
\end{figure}

Since $w(a) = 0$ for $a\ll a_{1/3}$, perturbations in the axion fluid behave like cold dark matter (CDM) perturbations until $a/a_{1/3}$ becomes large enough for the time dependence of the axion equation of state to become important. Accordingly, 
for modes entering the horizon sufficiently before $a_{1/3}$, both axion and CDM density perturbations grow according to the standard results for an RD era transitioning to MD: logarithmic growth during RD, transitioning to linear growth during MD. As in the standard cosmology, linear growth starts somewhat before MD as the influence of the matter perturbations on the corresponding metric modes becomes important before the background energy density of matter overtakes that of radiation. This initial logarithmic, then linear growth can be seen for CDM perturbations in the left panel of Fig.~\ref{fig:twofieldpert} for the two earliest-entering modes shown (brown and purple).  The right panel shows the  evolution of the corresponding metric perturbations.

\begin{figure}
\centering
\includegraphics[width=0.5\linewidth]{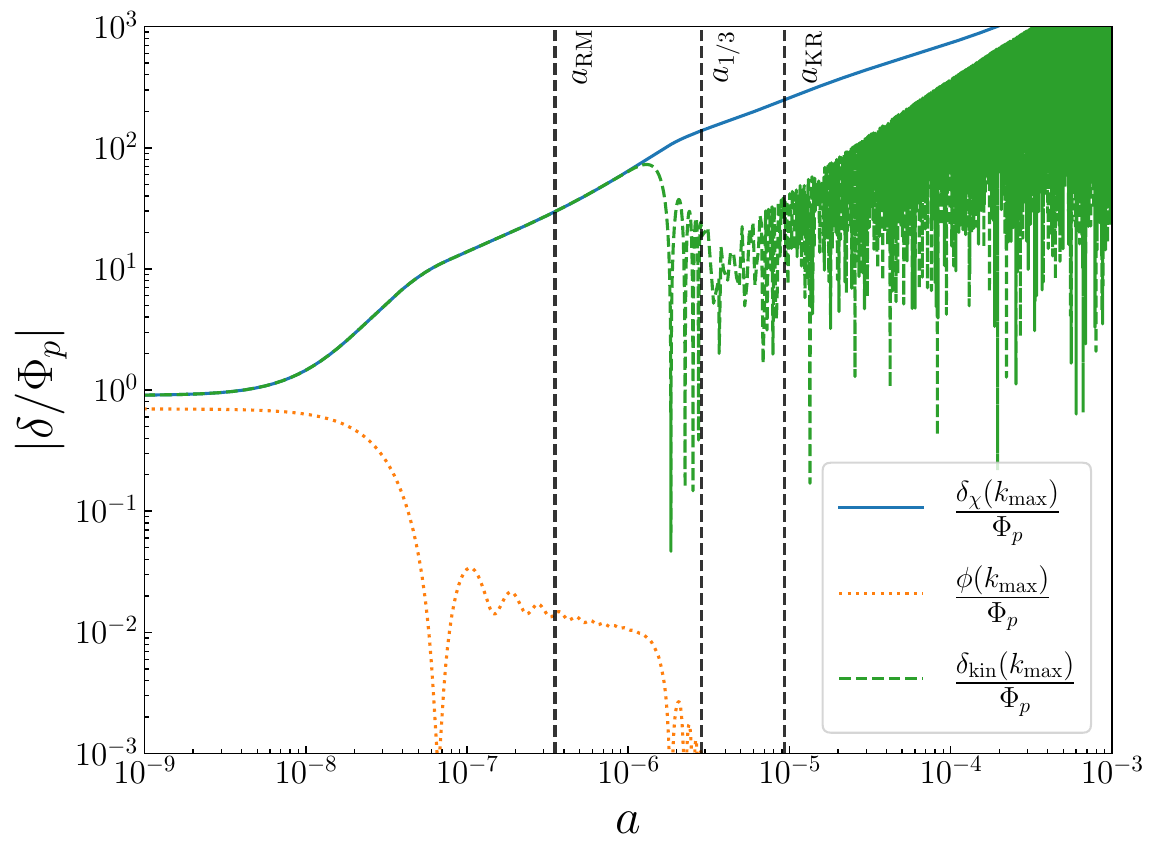}
\caption{Evolution of conformal Newtonian gauge dark matter ($\delta_{\chi}$), scalar metric $(\phi)$ and kinaton  perturbations $(\delta_{\rm kin})$ normalized to $\Phi_p$, the primordial curvature perturbation, for the mode $k_\mathrm{max}$ in the two-field model with  $T_\mathrm{KR} = 25 \eV $ and $T_\mathrm{RM} = 26.7~ T_{\rm KR}$ = 0.67 keV.}
\label{fig:twofieldpert1}
\end{figure}

The time-varying axion equation of state becomes important when the comoving wavelength of a perturbation becomes comparable to the comoving sound horizon, $k r_s \sim 1$.  Once an axion mode enters the sound horizon, the axion density perturbation starts to oscillate and the gravitational potential decays. This behavior can be seen in Fig.~\ref{fig:twofieldpert1}, where we show axion as well as CDM and metric perturbations for the mode $k_\mathrm{max}$ that experiences the maximal enhancement of the matter power spectrum relative to $\Lambda$CDM. For this mode, the axion density perturbation begins to oscillate around $a= 10^{-6}$.
Once the metric potential decays away,  DM density perturbations $\delta_\chi$ stagnate, while DM velocity perturbations $\Theta_\chi$ redshift as $1/a$.
As $c_s$ increases from zero, the modes with larger $k$ begin oscillating first. 
This can be seen by comparing the fall-off in $\phi$ around $a=10^{-6}$ for the two earliest-entering modes in the right panel of Fig.~\ref{fig:twofieldpert}, and the resulting end to linear growth for the two earliest-entering modes in the left panel.

For $a_{1/3}\lesssim a < a_{\rm KR}$, the axion drives an epoch of kination.
The evolution of cosmological perturbations during kination was established in \cite{Redmond:2018xty}.  During kination, the axion density perturbation $\delta_\theta$ oscillates rapidly after entering the horizon with an envelope that grows linearly with $a$, as also recently discussed in \cite{Eroncel:2025bcb}.  Sub-horizon DM modes grow linearly with $a$,  which can be understood as a result of the subhorizon drift of the DM particles after the decay of the gravitational potentials \cite{Redmond:2018xty}.

Given the limited observational windows for axion domination to occur post-BBN, the KD era in these cosmologies is brief: for the parameters shown in Figs.~\ref{fig:twofieldpert} and~\ref{fig:twofieldpert1}, KD lasts for less than a decade. Thus the background radiation energy density is not negligible throughout this epoch, and our CDM perturbations accordingly do not realize fully linear growth during the brief period of KD. What remains true is that all DM perturbation growth during our KD epoch is due to drift.
The DM mode that experiences the most growth during the kination epoch is thus the one that has the largest value of its velocity perturbation $\Theta_\chi$ going into the kination era. Since during MD $\Theta_\chi\propto a^{1/2}$ before entering the sound horizon and $\propto 1/a$ afterward, obtaining the largest value of $\Theta_\chi$ at $a_{1/3}$ is a trade-off between maximizing growth during MD and delaying the onset of oscillations. 

The growth obtained during MD depends weakly on the value of the model parameter $r_P$ through the time dependence of the sound horizon.
We find that during MD, $r_s\propto a^{7/2}$, with the coefficient of proportionality smaller for smaller values of $r_P$. 
Smaller values of $r_P$ imply that the sound horizon growth becomes important for a given value of $k$ at larger values of $a$, which stops perturbation growth later.  Thus values of $r_P$ that approach the enhanced-symmetry value $r_P =1$ give rise to a larger overall enhancement to the MPS; for instance, we find for $r_P = 1.1$ that the maximum enhancement of the MPS can be increased by a factor of a few. However, since the power-law dependence of $r_s$ on $a$ is independent of $r_P$, the location of the enhancement is insensitive to $r_P$.

In Fig.~\ref{fig:twofieldpert}, the mode that realizes the maximum enhancement with respect to $\Lambda$CDM is $k_\mathrm{max}$ (purple), which enters the horizon shortly before the onset of axion domination. As this figure shows, modes that enter later grow more quickly during the kination epoch, but do not catch up to modes that were subhorizon for the full duration of axion domination.
We find that the maximally-enhanced mode $k_\mathrm{max}$ is $O(100) k_{1/3}$, with the exact coefficient of proportionality depending weakly on the timing and duration of axion domination. Meanwhile, the magnitude of this enhancement scales approximately linearly with the length of axion domination; see the Appendix for more details. 

The falloff of the MPS enhancement at large $k$, seen in Fig.~\ref{fig:2fieldMPS}, is gradual, and is a combination of two effects.  First, the enhanced perturbation growth in the axion-dominated cosmology falls off with increasing $k$ as the stagnation of growth for modes with $kr_s>1$ during MD compounds with the less rapid growth such modes experience during KD. Second, the enhancement is being shown in relation to the growth experienced by the same mode in the standard cosmology, which  increases logarithmically with $k$.

Finally, the MPS is also modestly suppressed relative to $\Lambda$CDM for small $k$, corresponding to modes that enter the horizon late in the axion-dominated epoch and early in the subsequent RD era.  This effect was first noted by \cite{Sobotka:2024tat}, and arises because these modes enter the horizon later in the axion-dominated cosmology than they do in $\Lambda$CDM. One such mode is  $k_\mathrm{min}$ in Fig.~\ref{fig:twofieldpert} (orange).  For modes that enter late enough during axion domination, the limited growth they experience during the epoch of altered cosmological evolution is not sufficient to make up for the delayed onset of growth relative to $\Lambda$CDM.

\subsection{Log Potential}

\begin{figure}
\centering   
    \includegraphics[width=0.495\linewidth]{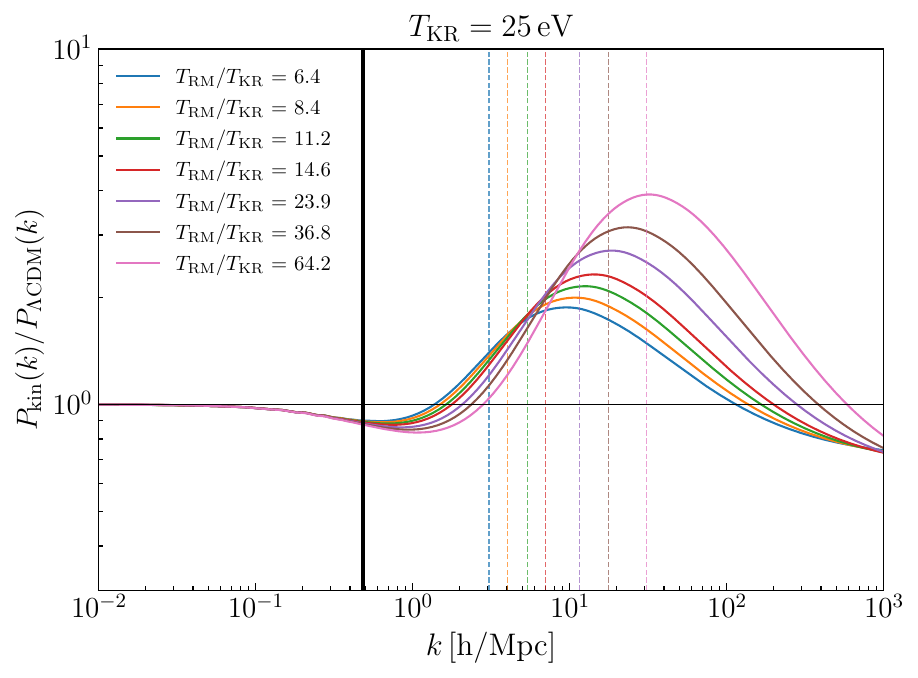}
    \includegraphics[width=0.495\linewidth]{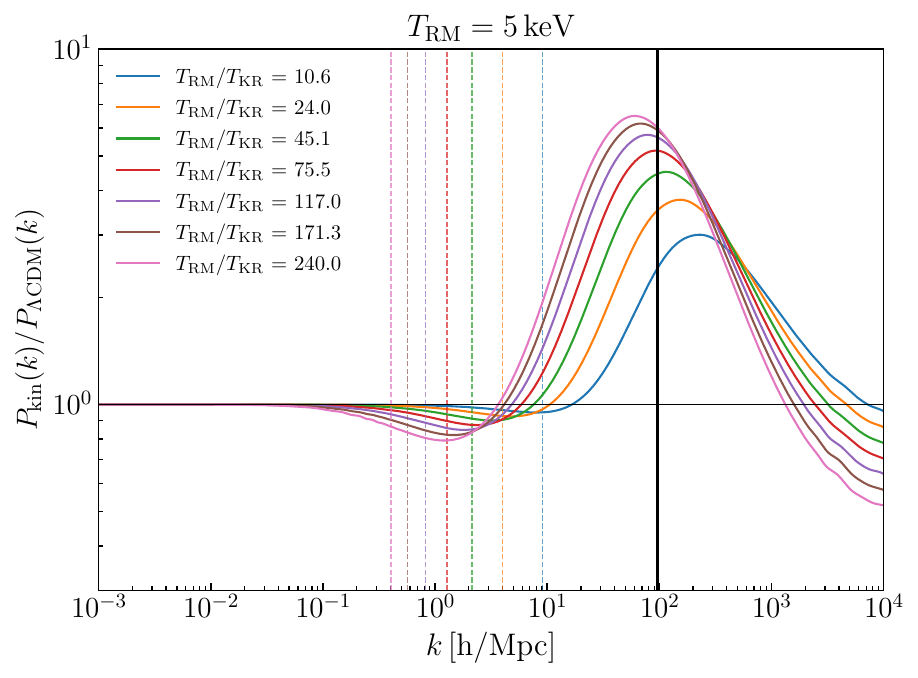}
\caption{\textbf{Left}: Ratio of the matter power spectrum  in the log model to the prediction of the standard cosmology ($T(k)  = P_{\rm kin}(k)/P_{\Lambda{\rm CDM}}(k)$) for fixed $T_{\rm KR} = 25 \eV$, with varying $T_{\rm RM}$; colors correspond to different total durations of axion domination as indicated in the legend. The black vertical line indicates $k_{\rm KR}$, while colored dotted vertical lines indicate $k_{\rm RM}$ for the correspondingly colored curve.  \textbf{Right}: Same as left except for  fixed $T_{\rm RM} = 5 \keV$ with varying endpoint $T_{\rm KR}$. Here the black line indicates $k_{\rm KR}$, while  colored dotted lines indicate $k_{\rm RM}$.}
\label{fig:logMPS}
\end{figure}

In the log potential model, the axion equation of state $w(a)$ evolves gradually  from $w\approx 0$ to $w=1/3$ during the initial ``matter''-dominated  phase of axion domination, then more rapidly to $w=1$.  
In Fig.~\ref{fig:logMPS} we show the resulting imprint on the MPS, again showing results that hold  $T_{\rm KR}$ fixed at 25 eV (left) and $T_{\rm RM}$ fixed at 5 keV (right).

For the limited duration of axion domination possible between BBN and CMB epochs, the axion equation of state in the log model is always meaningfully nonzero, so that axion perturbations always oscillate soon after horizon entry rather than grow.  Metric perturbations therefore decay during the ``matter''-dominated period (i.e., $a_{\rm RM}<a< a_{1/3}$) as well as during KD ($a_{1/3}< a< a_{\rm KR}$).  As a result, in the log model, the imprint of the altered expansion history on DM perturbations comes through the interplay of two effects: (i) the kick that these modes experience upon horizon entry while the potential is still nonzero, and (ii) the different expansion rate that changes the growth of DM perturbations after the potential decays.  Qualitatively, this model has some similarities to the vEDE scenario of Ref.~\cite{Sobotka:2024tat}, but the kination phase is here preceded by matter rather than dark energy domination, and the feature we obtain in the MPS is accordingly distinct.


\begin{figure}
\centering   \includegraphics[width=0.495\linewidth]{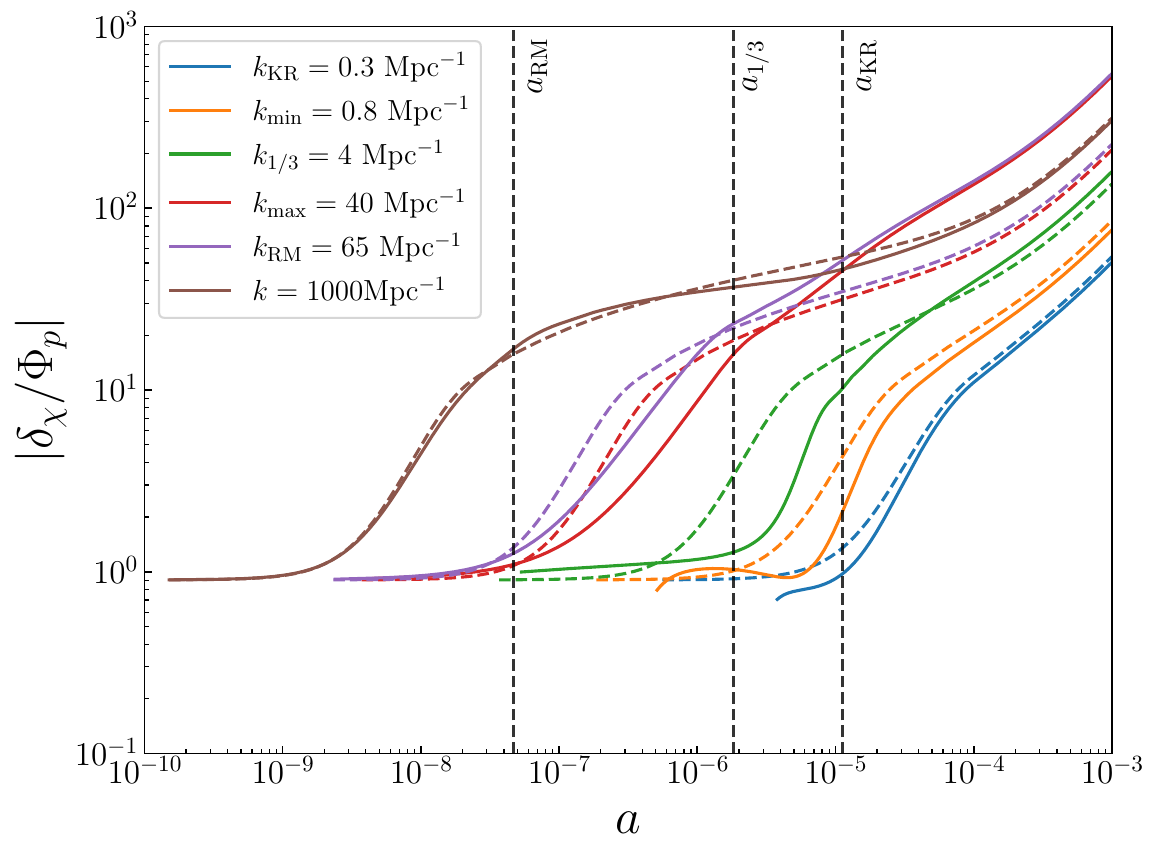}
\includegraphics[width=0.495\linewidth]{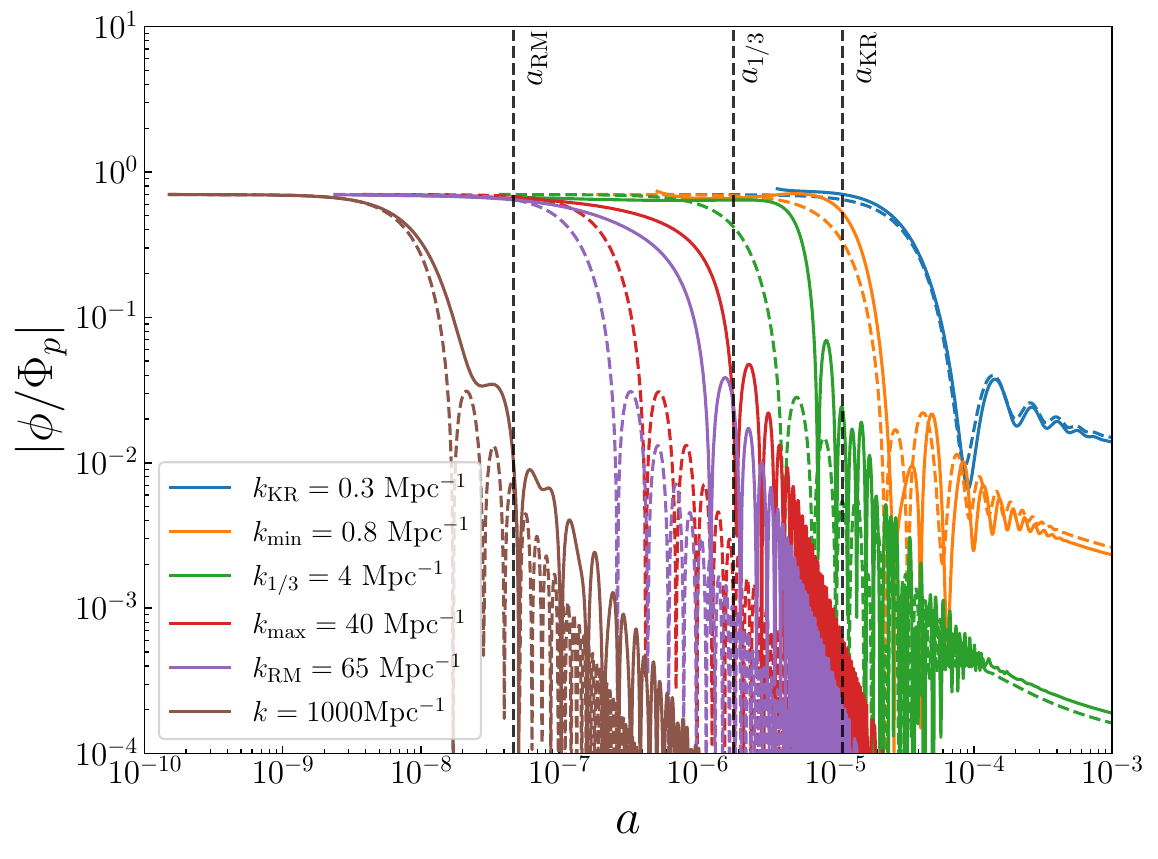}
\caption{
Evolution of conformal Newtonian gauge CDM perturbations $\delta_\chi$ (left) and metric perturbations $\phi$ (right) normalized to $\Phi_p$, the primordial curvature perturbation, with scale factor in a cosmology with axion domination in the log model (solid) and $\Lambda \rm{CDM}$ (dashed), for $T_\mathrm{RM} = 5 \keV$ and $T_\mathrm{KR} = 4 \times 10^{-3}~T_{\rm RM}$. Colors indicate different comoving wavenumbers $k$ as shown in the legend.
}
\label{fig:logpert1}
\end{figure}

\begin{figure}
\centering
\includegraphics[width=0.5\linewidth]{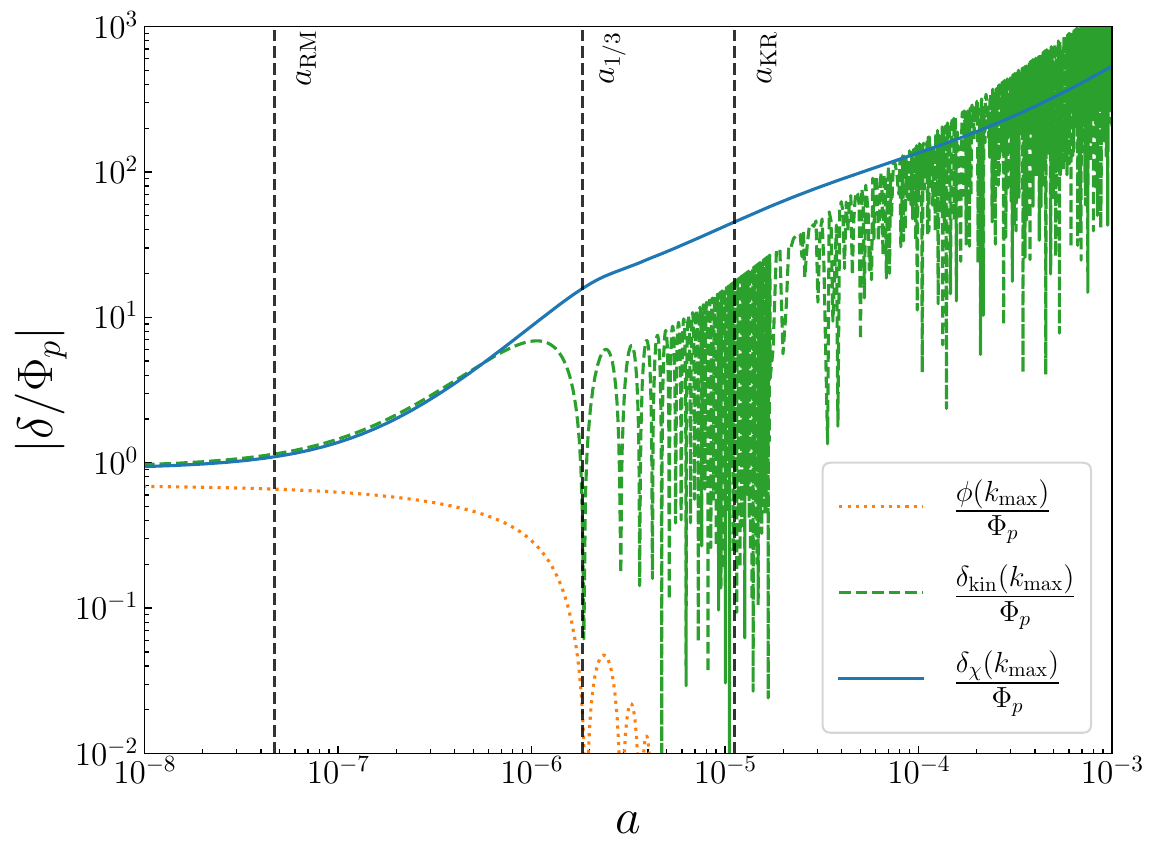}
\caption{Evolution of conformal Newtonian gauge matter perturbations ($\delta_{\chi}$), metric perturbations $(\phi)$, and axion perturbations $(\delta_\theta)$ normalized to $\Phi_p$, the primordial curvature perturbation, for the mode $k_\mathrm{max}$ in the log model for $T_\mathrm{RM} = 5 \keV$ and $T_\mathrm{KR} = 4 \times 10^{-3}~T_{\rm RM}$.
}
\label{fig:logpert}
\end{figure}


To discuss the two above effects more quantitatively, we show in Fig.~\ref{fig:logpert1} the evolution of several modes of $\delta_\chi$ and $\phi$ for a reference axion cosmology with  $T_\mathrm{RM} = 5 \keV$ and $T_\mathrm{KR} = 4 \times 10^{-3} T_{\rm RM} =$ 20 eV.   
During MD,
DM density perturbations $\delta_\chi$ begin growing at horizon entry ($k\tau \approx 1$) and cease growing once the axion begins to oscillate and the gravitational potentials start to damp ($\sqrt{w(a)} k\tau\sim 1$).  Since $w(a)\ll 1$, the onset of oscillations occurs later after horizon entry than it would during RD. Thus the initial CDM density perturbation growth is larger than the analogous growth from horizon entry during RD, but after oscillation begins, $\delta_\chi$ remains flat until $a\sim a_{1/3}$, as can be seen in the earliest-entering mode in Fig.~\ref{fig:logpert1} left (brown).  

For modes entering during MD, both the rate of DM growth and the duration over which the potentials source this growth decrease with decreasing $k$. Pre-oscillation, the effective source for the DM growth scales like $(k/(a\cH))^2 \phi$, and the potentials fall off more quickly with $a$ as $w$ increases toward $1/3$. Thus earlier-entering $\delta_\chi$ attain a larger amplitude {\em during MD} than do later-entering $\delta_\chi$, as can be seen in the comparison between $k_{\rm RM}$ (purple) and $k_{\mathrm{max}}$ (red) modes in Fig.~\ref{fig:logpert1}. However, modes that enter later during MD have a larger value of the velocity perturbation $\Theta_\chi$ going into the epoch of kination, which impacts the subsequent growth during kination. Modes entering earlier during MD (or, for shorter epochs of axion domination, earlier during the transition from RD to MD) thus experience both a longer period of stagnation during MD together with less kination-era growth.  This effect allows later-entering modes to overtake earlier-entering modes. 

For a given mode, the CDM velocity perturbation $\Theta_\chi$ reaches its maximum value at the onset of potential oscillations.  This maximum value depends on the magnitude of the kick received at horizon entry, which starts to fall off rapidly as $a$ approaches $a_{1/3}$. Therefore, the modes that experience the most CDM growth during kination are those where the potential began oscillating just before this fall-off, which translates into horizon entry roughly a decade prior to $a_{1/3}$.  The combination of initial growth at horizon entry during MD and linear growth during KD can give rise to a notable enhancement of the CDM density perturbation, though not as large as can be attained in the two-field model. The maximum growth is realized for comoving wavenumbers $k_\mathrm{max}= \mathcal{O}(10) k_{1/3}$, where $k_{1/3}$ is the wavenumber entering the horizon at $a_{1/3}$. 
We show the axion density perturbations for this mode, together with DM and metric modes, in Fig.~\ref{fig:logpert}, where the onset of axion oscillations just prior to $a_{1/3}$ is clearly visible.  The maximum growth achieved approximately scales with the duration of the kination epoch; see Appendix for more details.

As Fig.~\ref{fig:logMPS} shows, the MPS exhibits modest suppression relative to $\Lambda$CDM on both sides of the enhancement. The suppression at small $k$ again arises from delayed horizon entry.
For the log model, the MPS is also suppressed at large $k$, i.e., for modes that entered during RD and were sub-horizon throughout the entire epoch of axion domination. Here this suppression arises entirely through the changes to the Hubble rate in the modified cosmology, which changes the co-moving distances that the free-streaming DM particles can travel \cite{Redmond:2018xty}.  The growth of a DM density perturbation can be written formally in terms of the velocity and metric perturbations as
\begin{align}
\delta_\chi (a_f) - \delta_\chi(a_i) & = -\int_{a_i}^{a_f} da\left(\frac{\Theta_\chi}{a\cH}\right) + 3 \int_{a_i}^{a_f} da \,\phi' .
\end{align}
After the potentials have decayed to negligible values, the velocity perturbation evolves as $\Theta_\chi (a) = \Theta_0 /a$, and the growth of the density perturbation is entirely due to drift. 
The drift contribution to $\delta_\chi$ in the modified cosmology can be written as
\begin{equation}
\Delta \equiv - \int_{a_i}^{a_f} da\left(\frac{\Theta_\chi}{a\cH}\right) = -\Theta_0 \int_{a_i}^{a_f} \frac{da}{a^2\cH},
\end{equation}
where $a_i$ ($a_f$) should be chosen to be sufficiently small (large) that the contribution of the axion to the Hubble rate is negligible. 
In the log model, the time evolution of the conformal Hubble rate---i.e., matter-like, then kination-like---ensures that this drift contribution is smaller than its value in $\Lambda$CDM.  The resulting suppression in the MPS at high $k$ increases with the duration of the axion-dominated epoch, as is evident in Fig.~\ref{fig:logMPS}.

\section{Constraints}
\label{sec:constraints}

Late-time axion domination is stringently constrained by observations of both light element abundances and CMB anisotropies. In this section we build on the analysis of \cite{Co:2021lkc} to provide updated and model-specific estimates of the maximum duration of altered expansion that is compatible with both CMB and BBN. We discuss how the different equations of state for the two-field model and the log model result in slightly different constraints. In general the rotating axion can dominate for at most a decade, $T_{\rm RM}/T_{\rm KR} \lesssim 10$. 

The light element abundances resulting from BBN are sensitive to the onset of axion domination ($T_{\rm RM}$). As the additional energy density in the rotating axion becomes increasingly important relative to that of radiation, it raises the Hubble rate and thereby changes the timing of light element formation.  Meanwhile CMB anisotropies are primarily sensitive to the endpoint of axion domination ($T_{\rm KR}$), as the overall excellent agreement between CMB observations and the predictions of $\Lambda$CDM restricts altered perturbation evolution to the largest values of $k$ probed by the CMB. 

In order to establish the maximum extent of axion domination that is currently consistent with observation, we separately analyze (i) BBN constraints on the onset of axion domination ($T_{\rm RM}$) and (ii) CMB constraints on the endpoint of axion domination ($T_{\rm KR}$).
This procedure treats BBN and CMB constraints as largely uncorrelated, which is an excellent first approximation.  Specifically, in our BBN analysis we use Planck constraints on $\eta_B$ as a prior, but otherwise do not include CMB results.  In the CMB analysis we assume standard BBN relations between $\Omega_b h^2$ and $Y_P$.
These analysis choices are both conservative and self-consistent, as we discuss in detail below.

\subsection{Big Bang Nucleosynthesis}

To understand the impact of axion domination on BBN, we modify the publicly available code PRyMordial \cite{Burns:2023sgx} to include the time-dependent axion energy density using the equation of state $w(a)$ for both the log potential model and the two-field model with $r_P = 3$. 
We incorporate CMB constraints on $\eta_B$ by marginalizing over the constraint on  $\Omega_{b} h^2$ obtained from Planck while letting the helium mass fraction $Y_P$ float, i.e., without using the standard BBN consistency relation~\cite{Planck:2018vyg}. 
We also marginalize over twelve key nuclear rates impacting primordial deuterium and helium abundances (via the {\rm \texttt{smallnet} = True} flag in PRyMordial \cite{Burns:2023sgx}). We adopt the PRyMordial defaults in using $\tau_{n} = 878.4 {\pm 0.5}~ {\rm s}$ for the neutron lifetime and in employing nuclear rates from NACRE II. 
The marginalization is performed by sampling 1000 points randomly from the distributions of $\eta_{B}$, $\tau_{n}$ and the nuclear rates for a given $T_{\rm RM}$. This yields a distribution of predictions for the helium fraction $Y_P$ and the deuterium fraction $D/H$ with correlated uncertainty that is well-modeled as a Gaussian. The central value of the prediction, $Y_{P,{\rm BBN}}^0$ and $(D/H)_{{\rm BBN}}^0$ as well as the covariance matrix $\Sigma$ depend on $T_{\rm RM}$, and are evaluated from the results of the marginalization procedure.

We set constraints on $T_{\rm RM}$ while keeping the total duration of axion domination fixed at $T_{\rm RM}/T_{\rm KR} = 10$.  Since we test values of $T_{\mathrm{RM}}$ toward the end of BBN, this choice corresponds to assuming that the time variation in the axion equation of state is minimal through the range of scales probed by BBN. This is an excellent approximation for long epochs of axion domination ($T_{\rm RM}/T_{\rm KR} \gtrsim 10$), and is conservative for shorter epochs, especially for the log model.

\begin{figure}
\centering
    \includegraphics[width=0.32\linewidth]{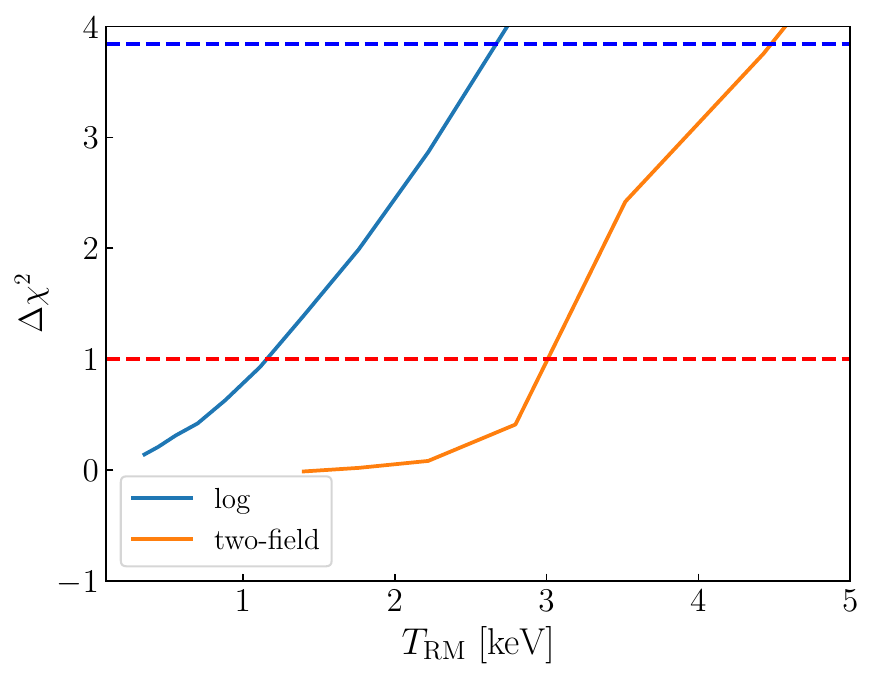}
    \includegraphics[width=0.32\linewidth]{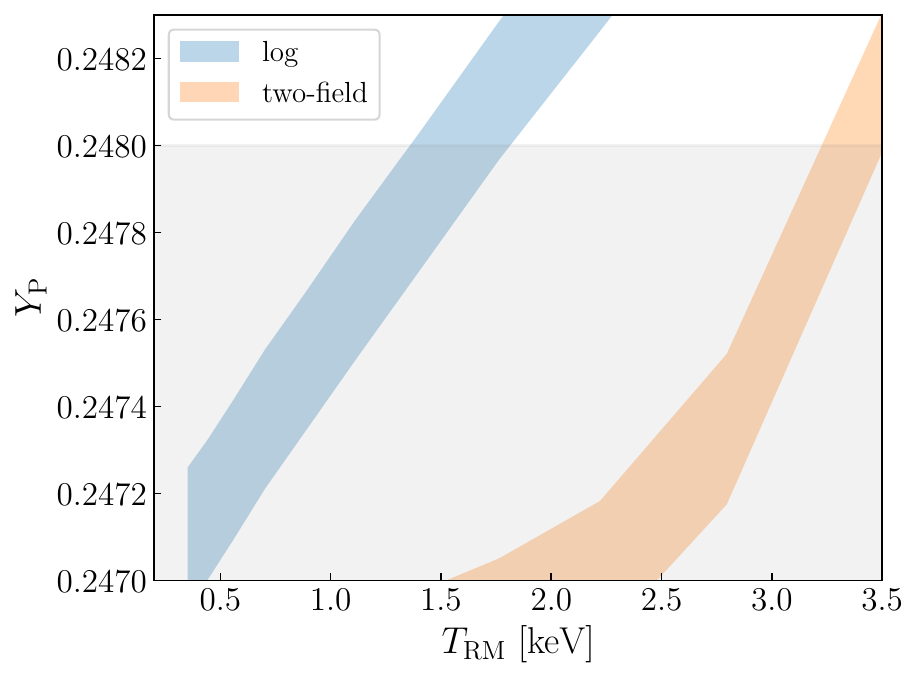}
    \includegraphics[width=0.32\linewidth]{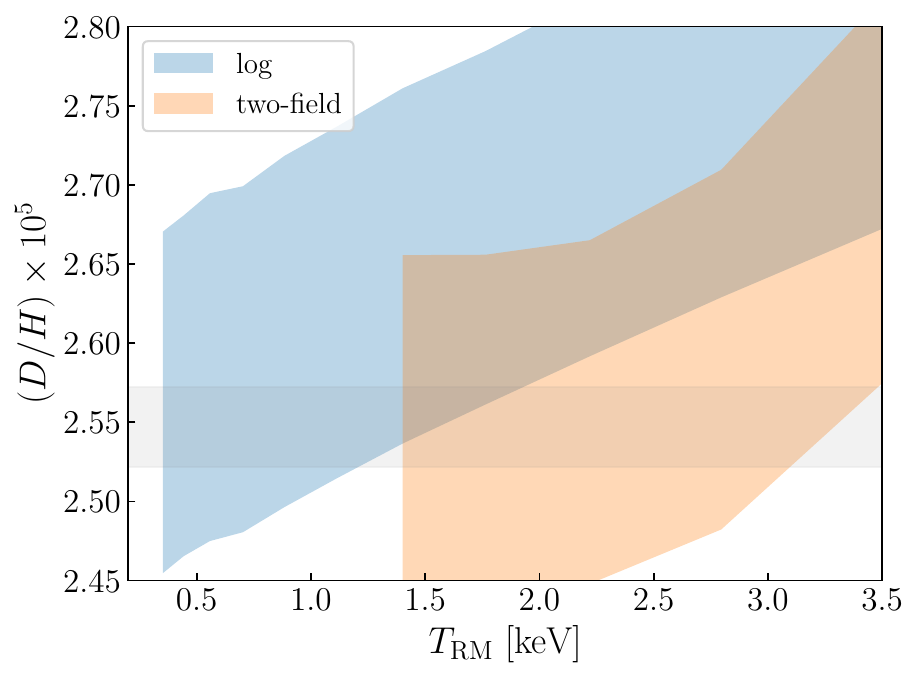}
\caption{BBN constraints on the log (blue) and two-field (orange) models. \textbf{Left:} $\Delta \chi^2$ between a given $T_{\rm RM}$ and $\Lambda{\rm CDM}$ Red (blue) dashed lines show $1\sigma (2\sigma)$ bounds. \textbf{Center:}  68\% confidence interval for primordial helium fraction $Y_{\rm P}$ as a function of $T_{\rm RM}$ in both models. The gray band indicates the observed value \cite{2020ApJ...896...77H,Aver:2020fon}. \textbf{Right:} 68\% confidence interval for primordial deuterium fraction $D/H$ as a function of $T_{\rm RM}$ in both models, with the observed value shown in gray \cite{Riemer-Sorensen:2017vxj, Cooke:2017cwo}.}
\label{fig:BBN}
\end{figure}

We define
\begin{align}
    \chi^2(T_{\rm RM},Y_P,D/H) = & \begin{pmatrix}
    Y_P - Y_{P,{\rm BBN}}^0,~D/H - (D/H)_{{\rm BBN}}^0
    \end{pmatrix} \Sigma^{-1}
    \begin{pmatrix}
    Y_P - Y_{P,{\rm BBN}}^0 \\ D/H - (D/H)_{{\rm BBN}}^0
    \end{pmatrix} \nonumber \\
    &+ \frac{(Y_P - Y_{P,{\rm obs}}^0 )^2}{\sigma_{Y_p}^2} + \frac{(D/H - (D/H)_{P,{\rm obs}}^0 )^2}{\sigma_{D/H}^2},
\end{align}
where $Y_{P,{\rm obs}}^0=  0.245$ and $(D/H)_{P,{\rm obs}}^0 \times 10^{5}=2.547$ are the central values of the measured light element abundances, and $\sigma_{Y_p}=0.003$ and $\sigma_{D/H}=0.025$ are their uncertainties
\cite{2020ApJ...896...77H,Aver:2020fon,Riemer-Sorensen:2017vxj, Cooke:2017cwo}.
We minimize $\chi^2$ with respect to $Y_P$ and $D/H$ to obtain $\chi^2_{\rm min}(T_{\rm RM})$, and define
\begin{align}
    \Delta \chi^2 = \chi^2_{\rm min}(T_{\rm RM}) - \chi^2_{\rm min}(T_{\rm RM}\rightarrow 0),
\end{align}
which is shown in Fig.~\ref{fig:BBN}. We require $\Delta \chi^2 < 1\,(3.84)$ for the $1\sigma(2\sigma)$ bound on $T_{\rm RM}$,
\begin{align}
    \label{eq:TRMbound}
    T_{\rm RM} &\leq 1.1~ (2.7) \keV \qquad {\rm~ for~log~model~at~1\sigma\, (2\sigma)},\\
    T_{\rm RM} &\leq 3 ~(4.5) \keV \qquad {\rm~ for~two\mbox{-}field~model~at~1\sigma\, (2\sigma)}.
\end{align}
The constraint on the log model is more stringent as the log model contributes more to the Hubble rate at $T> T_\mathrm{RM}$ than does the two-field model, which can be seen by comparing the energy densities of both models in Fig.~\ref{fig:rho}.
In Fig.~\ref{fig:BBN} we show the likelihood ratios for both models, as well as a comparison of the predicted helium and deuterium yields to the observed abundances.

It is worth observing that the suite of nuclear reaction rates adopted by PRIMAT underproduces deuterium in standard BBN for the CMB-preferred values of $\eta_B$ \cite{Pitrou:2020etk, Burns:2023sgx}. Since increasing the Hubble rate during BBN increases deuterium yield at a given $\eta_B$, using PRIMAT rates instead of NACRE II rates would substantially weaken the BBN constraint on~$T_{\rm RM}$.

\subsection{CMB power spectrum}
\label{sec:CMBcon}

To constrain the endpoint of axion domination, we perform a full likelihood analysis of CMB and BAO data using a modified version of \texttt{CLASS}~\cite{Blas:2011rf} coupled to \texttt{MontePython}~\cite{Brinckmann:2018cvx}. Our goal is to establish the longest duration of axion domination compatible with both BBN and CMB data.
Instead of performing a full combination of CMB and BBN constraints, we conservatively fix the onset temperature to around the largest values allowed by BBN within $1\sigma$, choosing $T_{\rm RM}=1\,\keV$ for the logarithmic model and $T_{\rm RM}=3\,\keV$ for the two-field model, and treat the transition temperature $T_{\rm KR}$ as a free parameter. Fixing $T_\mathrm{RM}$ at the upper bound of BBN compatibility  maximizes $T_{\rm RM}/T_{\rm KR}$ for a given $T_{\rm KR}$ and therefore maximizes the departure from $\Lambda$CDM (see Figs.~\ref{fig:2fieldMPS} and~\ref{fig:logMPS}), providing the most stringent constraint on $T_\mathrm{KR}$.

The BBN constraint on $T_{\rm RM}$ is dominated by deuterium observations (see Fig.~\ref{fig:BBN}). Since the dependence of $Y_P$ on $\eta_B$ is much flatter than that of deuterium, the main impact of the BBN analysis for the CMB is to enforce a relation between $\Omega_b$ and $Y_P$ very close to that of standard BBN. We therefore do not additionally marginalize over $Y_P$ in our CMB scans.

We assume spatial flatness and follow the Planck convention for neutrinos (two massless species and one massive with $m_\nu=0.06\,\eV$). Our sampled parameters are the six standard $\Lambda$CDM parameters, $ \{\omega_b,\,\omega_c,\,H_0,\,\ln(10^{10}A_s),\,n_s,\,\tau_{\rm reio}\},$ augmented by the axion parameter $T_{\rm KR}$. We fix $r_P=3$ in the two-field model. Flat priors are adopted for the $\Lambda$CDM parameters. For the transition temperature we use a log-flat prior and enforce the physical ordering $T_{\rm KR}<T_{\rm RM}$.

\begin{figure}
\centering
\includegraphics[width=\linewidth]{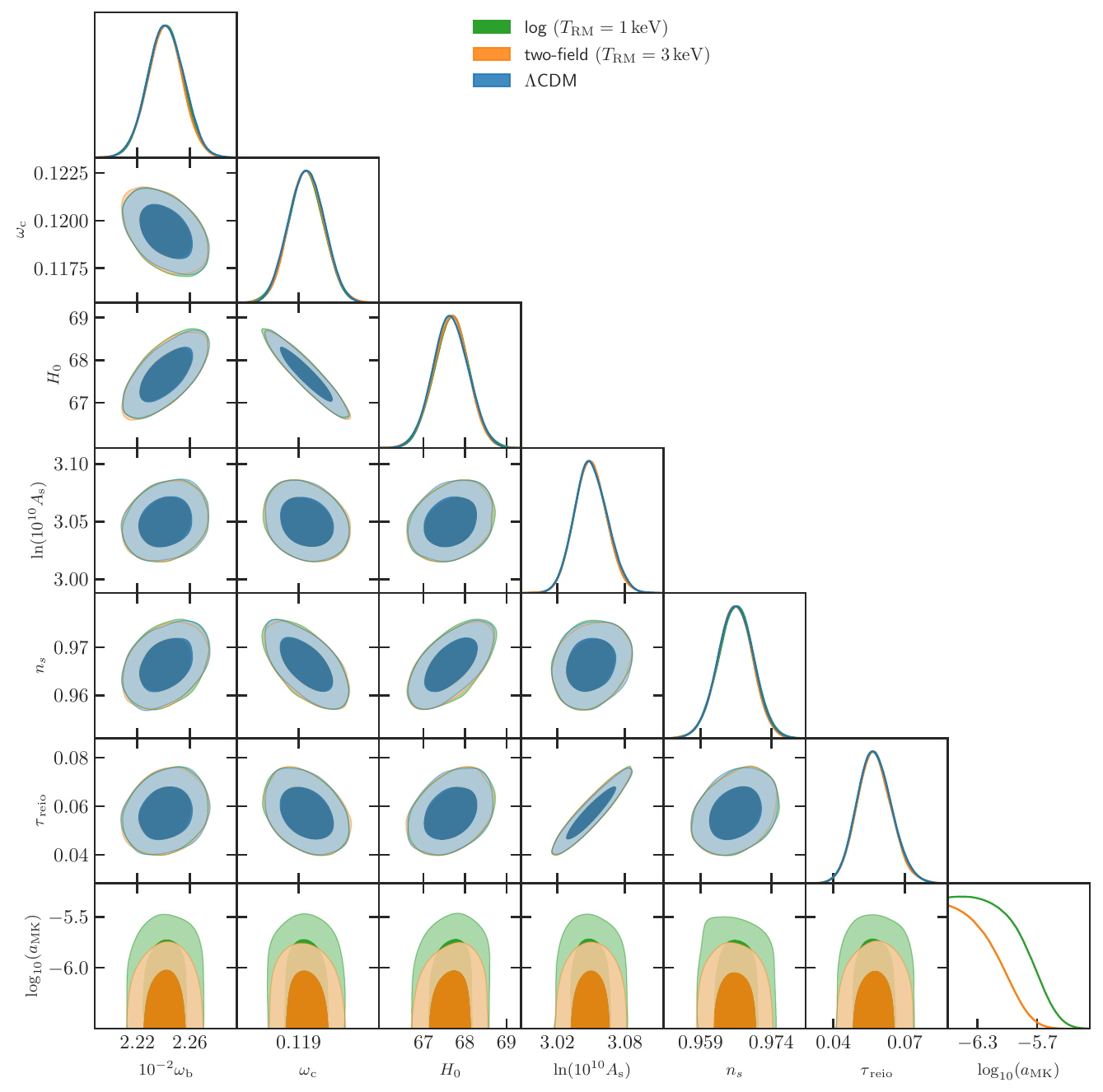}
\caption{Marginalized 2D posterior contours (68\% and 95\% C.L.) for the logarithmic (green), two-field (orange), and $\Lambda$CDM (blue) models, obtained from fits to CMB and BAO datasets discussed in the text.
}
\label{fig:MCMC}
\end{figure}

We use the following measurements:
\begin{itemize}
    \item \textbf{CMB:} Planck 2018 temperature and polarization spectra (low-$\ell$ TT, EE and high-$\ell$ TTTEEE) and the CMB lensing reconstruction~\cite{Planck:2018vyg}.
    \item \textbf{BAO:} Distance measurements from 6dFGS~\cite{Beutler:2011hx}, SDSS DR7 MGS~\cite{Ross:2014qpa}, and BOSS DR12 LOWZ/CMASS~\cite{BOSS:2016wmc}.
\end{itemize}
We run Markov Chain Monte Carlo with \texttt{MontePython}~\cite{Brinckmann:2018cvx} and declare convergence when the Gelman-Rubin criterion satisfies $|R-1|\leq 0.01$~\cite{Gelman:1992zz}.
Results are shown in Fig.~\ref{fig:MCMC}.

Similar to vEDE, the primary feature of late axion domination on CMB scales is the delayed horizon entry, relative to $\Lambda$CDM, for modes with $k\sim k_\mathrm{KR}$ \cite{Sobotka:2024tat}. The amplitude of this effect directly depends on the duration of axion domination. 
CMB and BAO data do not show any preference for this feature. At fixed $T_\mathrm{RM}$, raising $T_\mathrm{KR}$ both minimizes the amplitude of the axion feature and pushes it to higher $k$. 
We find the 95\% CL lower bounds on $T_\mathrm{KR}$ are
\begin{align}
    \label{eq:TKRbound}
    T_{\rm KR} &\geq \ \, 85 \eV \qquad {\rm~ for~log~model},\\
    T_{\rm KR} &\geq 102 \eV \qquad {\rm~ for~two\mbox{-}field~model}.
\end{align}
Since (i) the maximum value of $T_\mathrm{RM}$ allowed by BBN is larger for the two-field model than for the log model, and (ii) the equation of state for the two-field model predicts a larger integrated departure from $\Lambda$CDM at the background level, the constraint on $T_\mathrm{KR}$ for the two-field model is accordingly more stringent than in the log model.

Finally, as Fig.~\ref{fig:MCMC} makes clear, CMB constraints on $T_{\rm KR}$ are largely uncorrelated with the constraints on $\Lambda$CDM parameters, notably for our purposes $\Omega_b h^2$ and parameters such as $\tau$ and $n_s$ that affect the damping tail \cite{Co:2021lkc}. This indicates that the deformations the axion energy density makes to BBN predictions cannot simply be compensated with additional axion energy density at the recombination epoch, and validates the self-consistency of our~approach.

\section{Implications for the Small-Scale Matter Power Spectrum}
\label{sec:SSMPS}

\begin{figure}[t]
\centering   \includegraphics[width=0.95\linewidth]{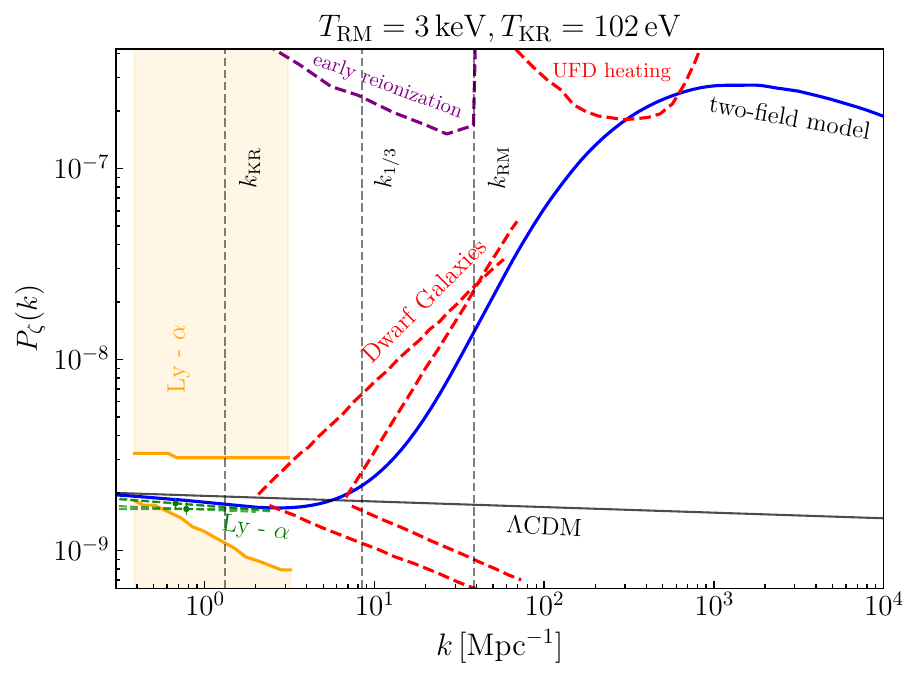}
\caption{Representative maximum enhancement to the MPS from late axion domination in the two-field model with $r_P=3$.  The black line is the primordial curvature power spectrum in $\Lambda\rm{CDM}$ given by $P_{\zeta,\Lambda\rm{CDM}} = A_{s}\left(\frac{k}{k_{*}}\right)^{n_{s} -1}$ where $A_{s} = 2.1 \times 10^{-9}$, $n_{s} = 0.96$, and  $k_{*} = 0.05 ~{\rm Mpc^{-1}}$ \cite{Planck:2018vyg}. The blue line shows $P_{\zeta,{\rm eff,kin}} = \frac{P_{\rm kin}(k)}{P_{\rm \Lambda{\rm CDM}}(k)} P_{\zeta,\Lambda{\rm CDM}}$ where $P_{\rm kin}$/$P_{\Lambda \rm{CDM}}$ is the ratio of the  MPS in kination  cosmology to that in $\Lambda{\rm CDM}$. 
The orange shaded region shows model-independent bounds on the MPS from Lyman-$\alpha$ forest observations \cite{Bird:2010mp}. Other constraints derived under specific assumptions about the shape of the small-scale MPS are shown in dashed lines: eBOSS Lyman-$\alpha$ observations of the slope and amplitude of the MPS at pivot scales $k_P \approx  1\,h~{\rm Mpc^{-1}}$  (green) \cite{Fernandez:2023grg,eBOSS:2018qyj}; properties of Milky Way satellites (red) \cite{Dekker:2024nkb}; dynamical heating of ultra-faint dwarf galaxies  (red) \cite{Graham:2024hah}; and early reionization (purple) \cite{Qin:2025ymc}. 
}
\label{fig:fiducial2field}
\end{figure}

\begin{figure}[t]
\centering   \includegraphics[width=0.95\linewidth]{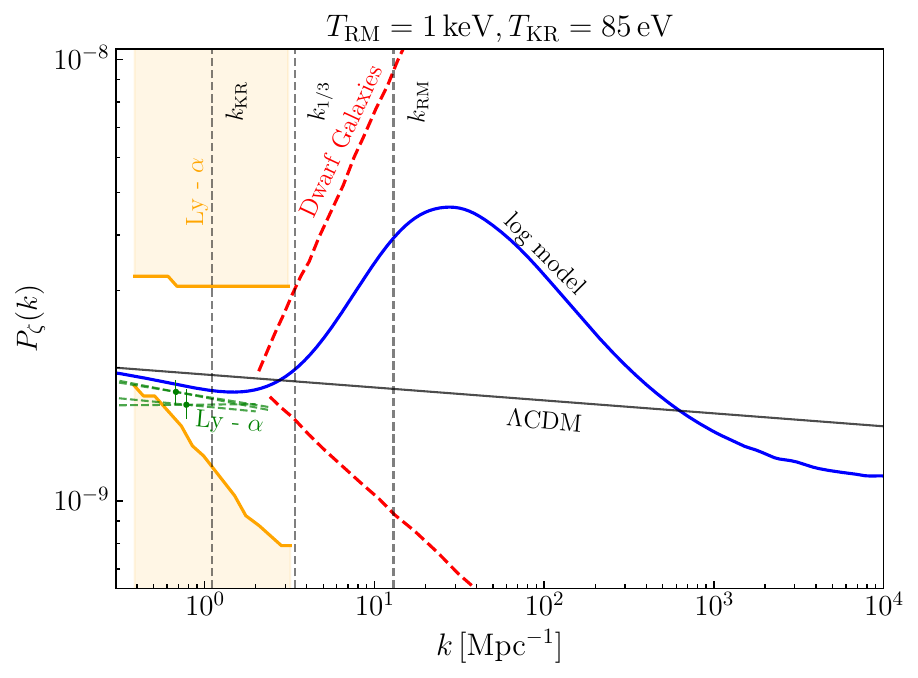}
\caption{Representative maximum enhancement to the MPS from late axion domination in the log model. 
The black line is the primordial curvature power spectrum in $\Lambda\rm{CDM}$ given by $P_{\zeta,\Lambda\rm{CDM}} = A_{s}\left(\frac{k}{k_{*}}\right)^{n_{s} -1}$ where $A_{s} = 2.1 \times 10^{-9}$, $n_{s} = 0.96$, and  $k_{*} = 0.05 ~{\rm Mpc^{-1}}$ \cite{Planck:2018vyg}. The blue line shows $P_{\zeta,{\rm eff,kin}} = \frac{P_{\rm kin}(k)}{P_{\rm \Lambda{\rm CDM}}(k)} P_{\zeta,\Lambda{\rm CDM}}$ where $P_{\rm kin}$/$P_{\Lambda \rm{CDM}}$ is the ratio of the  MPS in kination  cosmology to that in $\Lambda{\rm CDM}$, assuming Planck 2018 best-fit parameters.  
The orange shaded region shows model-independent bounds on the MPS from Lyman-$\alpha$ forest observations \cite{Bird:2010mp}. Other constraints derived under specific assumptions about the shape of the small-scale MPS are shown in dashed lines: eBOSS Lyman-$\alpha$ observations of the slope and amplitude of the MPS at pivot scales $k_P \approx  1\,h~{\rm Mpc^{-1}}$  (green) \cite{Fernandez:2023grg,eBOSS:2018qyj}; and properties of of Milky Way satellites (red) \cite{Dekker:2024nkb}.
}
\label{fig:fiduciallog}
\end{figure}

We show our results for the small-scale MPS for the two-field and log potential models in Figs.~\ref{fig:fiducial2field} and~\ref{fig:fiduciallog} respectively.  For these figures we consider the maximum duration of axion domination allowed by CMB anisotropies and BBN; specifically, we fix $T_\mathrm{RM} = 1$ keV for the log model and 3 keV for the two-field model, and set $T_\mathrm{KR}$ to the  values
that saturate the 95\% CL bounds obtained from CMB observations in the previous section.
The power spectra shown in these figures therefore realize the maximum achievable enhancement within each model \footnote{Somewhat larger enhancements can be realized in the two-field model for smaller values of $r_P$.}.
However, the overall detectability can be enhanced for smaller $T_{\rm RM}$, as decreasing $T_\mathrm{RM}$ at fixed $T_{\rm KR}$ shifts the peak  to lower $k$ (albeit with a smaller amplitude, see Figs.~\ref{fig:2fieldMPS} and \ref{fig:logMPS}). Decreasing $T_\mathrm{RM}$ also allows for smaller values of $T_\mathrm{KR}$, which gives rise to a milder small-$k$ suppression that can extend to lower $k$.

In these figures we show the primordial curvature power spectrum in $\Lambda\rm{CDM}$ in black, given by $P_{\zeta,\Lambda\rm{CDM}} = A_{s}\left(\frac{k}{k_{*}}\right)^{n_{s} -1}$ where we take $A_{s} = 2.1 \times 10^{-9}$, $n_{s} = 0.96$, and the pivot scale $k_{*} = 0.05 ~{\rm Mpc^{-1}}$ \cite{Planck:2018vyg}. The predictions of the {\em effective} curvature power spectrum for late axion domination are shown in blue as $P_{\zeta,{\rm eff, kin}} = \frac{P_{\rm kin}(k)}{P_{\rm \Lambda{\rm CDM}}(k)} P_{\zeta,\Lambda{\rm CDM}}$ where $P_{\rm kin}$/$P_{\Lambda \rm{CDM}}$ is the ratio of the  MPS in kination  cosmology to that in $\Lambda{\rm CDM}$, assuming Planck 2018 best-fit parameters. We emphasize {\em effective} because even though we illustrate our results in terms of the primordial curvature power spectrum, the enhancement of the matter power spectrum happens through subhorizon evolution and not during inflation.

There are numerous probes of the small-scale power spectrum. However, the majority of these probes rely on observations of structures in  mildly to deeply non-linear regimes, so the implications of the resulting constraints for scenarios with late axion domination are in general not entirely straight-forward to assess.  In Figs.~\ref{fig:fiducial2field} and~\ref{fig:fiduciallog} we show constraints that are obtained assuming a different functional form of the linear MPS with dashed lines.

\paragraph{Lyman-$\alpha$ forest.} The absorption spectra of distant quasars probe  the distribution of baryonic gas in the intergalactic medium on scales up to tens of Mpc, making it a powerful probe of the matter power spectrum in the mildly non-linear regime. 
We show constraints on the MPS from a minimally parametric reconstruction due to \cite{Bird:2010mp}, using observations of quasar spectra from the Sloan Digital Sky Survey (SDSS).
These constraints are shown as the orange shaded region in Figs.~\ref{fig:fiducial2field} and~\ref{fig:fiduciallog}.

Analyses based on the larger data set provided by the extended Baryon Oscillation Spectroscopic Survey (eBOSS) quasar spectra yield more powerful constraints on the power spectrum.  
The authors in \cite{eBOSS:2018qyj,Fernandez:2023grg} parametrize the small-scale power spectrum with an amplitude $A_{P}$ and power law with slope $n_{P}$ at pivot scales of $k_P = 0.78/$Mpc  and $k_P = 1 h/{\rm Mpc} \approx 0.67/$Mpc, respectively. Both analyses find a suppression of the matter power spectrum compared to $\Lambda{\rm CDM}$, with some disagreement about the slope. These results are shown in the dashed green lines  in Figs.~\ref{fig:fiducial2field} and~\ref{fig:fiduciallog}.  While our axion kination models do result in suppressed power on the relevant scales, and thus on these scales improve agreement with observations, the results should be interpreted with caution. 
Although the fiducial spectra shown in the figures are nearly monotonic over the range of scales nominally probed by  eBOSS quasar spectra, it is worth establishing to what extent the enhanced power on somewhat smaller scales impacts the predicted flux spectra. For shorter epochs of axion domination than shown in Figs.~\ref{fig:fiducial2field} and~\ref{fig:fiduciallog}, the power spectrum will not be monotonic on Lyman-$\alpha$ scales.
It is thus necessary to perform a dedicated analysis of the Lyman-$\alpha$ data with the power spectrum predicted by axion kination cosmology to make more reliable quantitative statements.

\paragraph{Dwarf galaxies.} 
An enhancement of the matter power spectrum above a few Mpc$^{-1}$ affects the concentration as well as the abundance of dwarf galaxies, which can be a powerful probe of the small-scale MPS \cite{Esteban:2023xpk}.  In \cite{Dekker:2024nkb} the authors parametrize the enhancement (suppression) to the matter power spectrum as a blue (red) tilt above a reference scale $k_{p}$ and perform detailed modeling of galaxy formation  to make predictions for the resulting dwarf galaxy properties.  Comparing the results to a sample of Milky Way dwarf satellites allows them to derive constraints on the enhancement (suppression) of the matter power spectrum. Selected results from \cite{Dekker:2024nkb} are shown by red dashed lines in Figs.~\ref{fig:fiducial2field} and~\ref{fig:fiduciallog}, specifically, the limits on the power spectrum assuming a reference scale $k_{p} = 3 h^{-1} ~{\rm Mpc}$ and also $k_{p} = 10 h^{-1} ~{\rm Mpc}$  for the two-field model.  

 These constraints are likely somewhat overstated for our non-monotonic features, as the process of non-linear structure formation transfers power from smaller to larger scales, as recently demonstrated in the context of small-scale enhancements to the MPS in \cite{Nadler:2025crd}.

Constraints on the abundances and concentrations of low mass halos can be obtained by studying perturbations to strongly-lensed systems \cite{Gilman:2021gkj}. This analysis similarly considers a broken power law form for the primordial power spectrum, and finds constraints on the small-scale MPS that are broadly comparable to those from dwarf galaxy concentrations.

In \cite{Graham:2024hah} the authors consider the heating effects on stars in ultra-faint dwarf galaxies from dynamical friction due to a population of  concentrated dark subhalos, which enables them to constrain enhancements to the matter power spectrum on comoving wavenumbers  $k\gtrsim 10/{\rm Mpc}$. The excluded parameter space is shown in dashed red in Fig.~\ref{fig:fiducial2field}.  These constraints are obtained assuming a narrow enhancement in the primordial power spectrum at a given $k$, and are shown as a reference.
While the log model does not realize the kinds of enhancements to which dwarf heating is sensitive, the two-field model can, and indeed dwarf heating is likely already testing the maximal spectrum shown in Fig.~\ref{fig:fiducial2field}. A dedicated analysis of sensivities to this broad feature is thus particularly interesting for future work.

\paragraph{Early star formation.} 
If early-forming dark matter halos are large enough to enable atomic cooling, they can host star formation at earlier redshifts than in standard cosmology.
Ionizing radiation from these early stars and galaxies causes the hydrogen in the universe to re-ionize earlier than bounds inferred from CMB and  Lyman-$\alpha$ observations.  We show the exclusions obtained in \cite{Qin:2025ymc} in dashed purple in Fig.~\ref{fig:fiducial2field}. Similar to the bounds on ultra-faint dwarf heating, these limits are obtained assuming a narrow enhancement in the primordial power spectrum at a given $k$.
Our fiducial scenario does not realize large enough enhancements on the relevant scales to make these limits obviously constraining.

A more modest enhancement on similar scales can give rise to an enhanced star formation rate at large redshifts $(z \gtrsim 10)$. This early star formation can
lead to enhanced contributions to the background intensity of ultra-violet light. The  UV luminosity function observed by the Hubble Space Telescope then puts an upper bound on the early star formation rate and correspondingly the enhancement to the matter power spectrum on scales $0.5 \,\mathrm{Mpc}^{-1}\lesssim k \lesssim \mathrm{few} \,\mathrm{Mpc}^{-1}$ \cite{Sabti:2021unj,Sabti:2023xwo}. These scales can receive moderate enhancements in the fiducial axion cosmologies shown in Figs.~\ref{fig:fiducial2field} and~\ref{fig:fiduciallog}. The power on these scales is particularly sensitive to the values of $T_{\rm RM}$ and $T_{\rm KR}$ and may be increased or decreased while remaining consistent with the BBN and CMB constraints. Detailed predictions from late axion domination for early star formation and the resulting UV luminosity functions are thus an interesting topic for further investigation.

\paragraph{CMB spectral distortions.} CMB $\mu$-distortions are powerful probes of enhancements to the small-scale primordial curvature power spectrum. However, late axion domination realizes enhancements to matter perturbations through sub-horizon evolution.  In these cosmologies, the scale-dependent growth experienced by photon perturbations differs in general from that of CDM perturbations, and typically is substantially smaller.  Using a tight-coupling approximation to the heating rate \cite{Chluba:2015bqa}, we estimate that the $\mu$-distortions realized in the two-field model are no more than an order of magnitude larger than the $\Lambda$CDM predictions, and for the log model can even be suppressed.  Thus late axion domination can make interesting predictions for instruments such as PIXIE \cite{Kogut:2011xw} that would be sensitive to the expected signal from $\Lambda$CDM, but there are presently no constraints. 

Another class of models that can yield an enhanced matter power spectrum on small scales includes axion DM from
the misalignment mechanism with inflationary quantum fluctuations in the initial misalignment angle~\cite{Linde:1984ti}
or DM produced from inflationary quantum fluctuations \cite{Chung:2004nh,Chung:2013rda,Graham:2015rva, Alonso-Alvarez:2018tus,Kolb:2023ydq}. In these scenarios the enhanced small-scale DM perturbations are isocurvature, which also gives rise to a relatively smaller spectral distortion signature for the same MPS enhancement~\cite{Chluba:2013dna}. 
Since the origin and evolution of the enhanced perturbations in late axion domination cosmologies are different from scenarios with primordial isocurvature enhancements \cite{Buckley:2025zgh}, we expect the spectral distortion signatures to differ as well.  Thus we emphasize that spectral distortions will be a powerful discriminator of the origin of any enhancement seen in the small-scale MPS.

\section{Summary and Discussion}
\label{sec:summary}

In this paper, we considered a rotating axion field in the early universe, which can explain the baryon asymmetry of the universe. This rotation enables the axion to realize a period of matter, then kination domination between Big Bang Nucleosynthesis and recombination. We found that the evolution of DM perturbations through this modified expansion history gives rise to substantial modifications to the matter power spectrum on the scales 
$1/\mathrm{Mpc}\lesssim k \lesssim 10^3/$Mpc. Imposing BBN and CMB constraints, we find the log model may give a modest enhancement of up to a factor of a few with a bumpy feature, while the two-field model may exhibit an  $O(100)$ enhancement with a plateau-like feature, as shown in Figs.~\ref{fig:fiducial2field} and~\ref{fig:fiduciallog}. 
Intriguingly, these models can realize the modest suppression of power on $1/$Mpc scales suggested  by recent Lyman-$\alpha$ analyses \cite{Fernandez:2023grg,Rogers:2023upm}, while still predicting substantial enhancements at smaller scales.
In those figures, we chose the temperatures marking the start and end of axion domination to obtain the maximal enhancement at the peak. By choosing other sets of temperatures, we may shift the scale that receives the most enhancement to smaller $k$, while decreasing the overall amplitude of the MPS. Additionally, the amplitude of the feature predicted in the two-field model may be enhanced by a factor of a few by taking the model parameter $r_P\rightarrow 1$.

Enhancements to the MPS predict a larger population of earlier-forming and thus more concentrated dark matter structures on the corresponding mass scales.  
These axion domination models can therefore provide a compelling theoretical target for the numerous observational probes of DM substructure that are expected to substantially advance in sensitivity with the advent of JWST and new optical surveys. 
Interestingly, there exist measurements that can be interpreted as hints of enhanced structure formation, such as the numerous extremely luminous objects (little red dots) at high redshifts $(z \gtrsim 10)$ seen in JWST.  Some of these sources have been confirmed to be accreting supermassive black holes (SMBHs)~\cite{2024NatAs...8..126B, 2023ApJ...957L...7K,2023ApJ...953L..29L,2024Natur.628...57F,2023MNRAS.525.1353J,2024arXiv241018763N}, while others have been confirmed as early star-forming galaxies  at redshifts of $z\approx 14$~\cite{2024Natur.633..318C}. While JWST observations may be compatible with standard $\Lambda {\rm CDM}$ cosmology~\cite{2023ApJ...955L..35S,Feldmann:2024kwz,Ellis:2025xju}, we look forward to further observations and analyses from JWST that will constrain or find evidence for enhancement of MPS.  Separately, certain strongly-lensed systems may have subhalos with higher concentration than expected in $\Lambda {\rm CDM}$  \cite{Hezaveh:2016ltk,Despali:2024ihn,Enzi:2024ygw,Sengul:2021lxe,Lange:2024pef,Vegetti:2023mgp}, although the limited number of these systems, together with the challenges of detecting sub-halos \cite{Nightingale:2022bhh, Ephremidze:2025mqg}, makes it premature to definitively ascribe these high concentrations to new particle physics.

In \cite{Qin:2025ymc}, the authors point out that the substantial abundance of high-redshift SMBHs revealed by  JWST  can be explained with a small-scale enhancement to the MPS around $k=100/$Mpc that enables the early formation of large BH seeds  from direct collapse.  Our two-field model realizes an enhancement to the MPS with a similar size and location to the primordial plateau considered in \cite{Qin:2025ymc}, and we thus expect early formation of direct collapse BHs to be a compelling consequence of the two-field model.

On dwarf galaxy scales, high-redshift measurements of the small-scale MPS are intricately entwined with measurements of early star formation. Late axion domination generally realizes modest enhancements on these scales, and understanding the resulting predictions for non-linear structure formation in detail will be of interest as both measurements and understanding of star formation continue to improve. On this front, future 21-cm observations will provide a powerful additional probe of star formation as well as the small-scale MPS \cite{Munoz:2019hjh}.  
Low-redshift measurements of MW satellites, taken at face value, already yield constraints that are interestingly close to the maximal predictions of the two-field model \cite{Dekker:2024nkb}.  We thus expect that the fiducial two-field model may be an interesting target for future such analyses, though work would be required to establish the predictions for dwarf galaxy populations arising from our non-monotonic features in the MPS. 

The largest enhancements we are able to realize occur for wavenumbers corresponding to halos too small to host stars. Here as well, future work developing the consequences of late axion domination for nonlinear structures in the low-redshift universe will be important for making predictions for observables such as dwarf heating and stellar stream disruption that are sensitive to  the gravitational impact of concentrated dark subhalos. In particular, ultra-faint dwarf heating \cite{Graham:2024hah} stands out as a potentially powerful test of the largest plateau~features. 

Small-scale structure is well-established as a powerful probe of the particle physics of DM. Here we emphasize that the small-scale MPS also provides invaluable information about the expansion history of the universe in the epoch between BBN and reionization. While the thermal history of the universe in this era  is tightly constrained, there is still room for well-motivated particle physics models to yield large deviations from the standard cosmology. Indeed, in the late axion domination models presented here, measurements of small-scale structure could ultimately provide our first observational window onto baryogenesis.

\acknowledgments

We acknowledge helpful conversations with Hengameh Bagherian, Simeon Bird, Priyesh Chakraborty, Chandrika Chandrashekhar, Adrienne Erickcek, Siu Cheung Lam and Sai Chaitanya Tadepalli. We especially thank Adrienne Erickcek for comments on the manuscript. This work was supported by the GRASP initiative at Harvard University (AG); by the DOE grants DE-SC002561 (RC), DE-SC0010008 (NF), DE-SC0009924 (KH), and DE-SC0015655 (JS); and by World Premier International Research Center Initiative (WPI), MEXT, Japan (Kavli IPMU) (KH).
R.C.~and N.F.~thank the Aspen Center for Physics, which is supported by the National Science Foundation grant PHY-2210452, and the a grant from the Simons Foundation (1161654, Troyer).
For facilitating portions of this research, R.C. wishes to acknowledge the Center for Theoretical Underground Physics and Related Areas (CETUP*), the Institute for Underground Science at Sanford Underground Research Facility (SURF), and the South Dakota Science and Technology Authority for hospitality and financial support, as well as for providing a stimulating environment when this work was finalized. J.S.~thanks the Kavli Institute for Theoretical Physics, supported by the NSF grant PHY-2309135, for hospitality during the performance of this work.

\appendix

\section{Scaling properties of axion-induced power spectrum features}

In this Appendix we provide additional information about how the properties of the features in the MPS depend on the timing and duration of axion kination domination, using the example spectra of Sec.~\ref{sec:pertgrowth}, shown in Figs.~\ref{fig:2fieldMPS} and~\ref{fig:logMPS}.  In general, the lack of large separation of scales and the non-polynomial equations of state exhibited in these models give rise to detailed parametric dependence that is not well-described by simple power laws. In this section we present
 approximate power law fits to establish how the location $k_\mathrm{max}$ and size $(P_{\rm kin}/ P_{\Lambda {\rm CDM}})|_{\rm max}$ of the maximum enhancement scale with the duration of axion kination domination. These approximate power laws serve as useful rules of thumb within the parameter ranges shown, which bracket the parameters relevant for observably allowed spectra.

\begin{figure}
\centering   
    \includegraphics[width=0.495\linewidth]{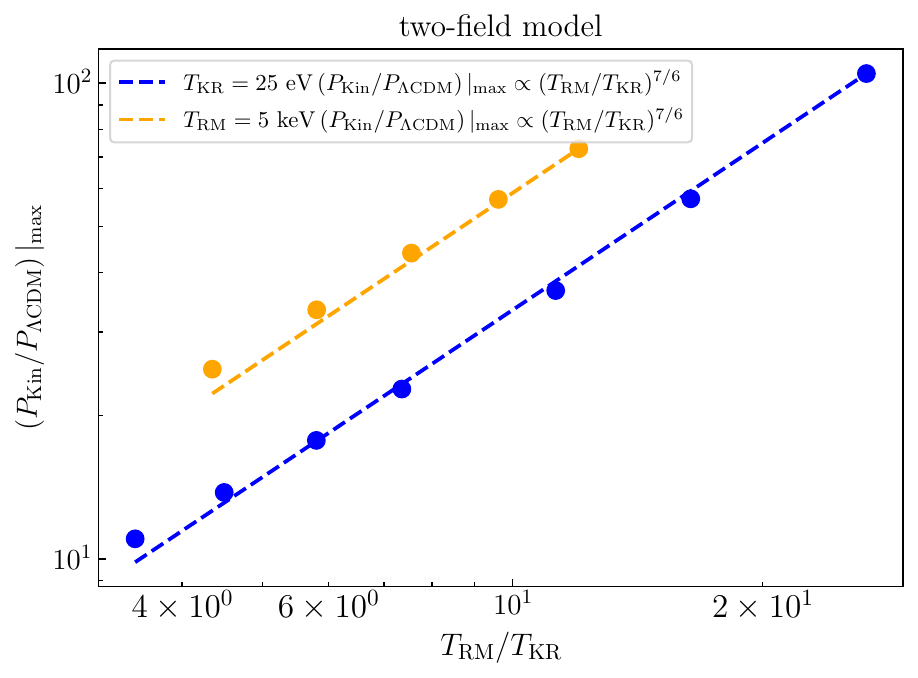}
    \includegraphics[width=0.495\linewidth]{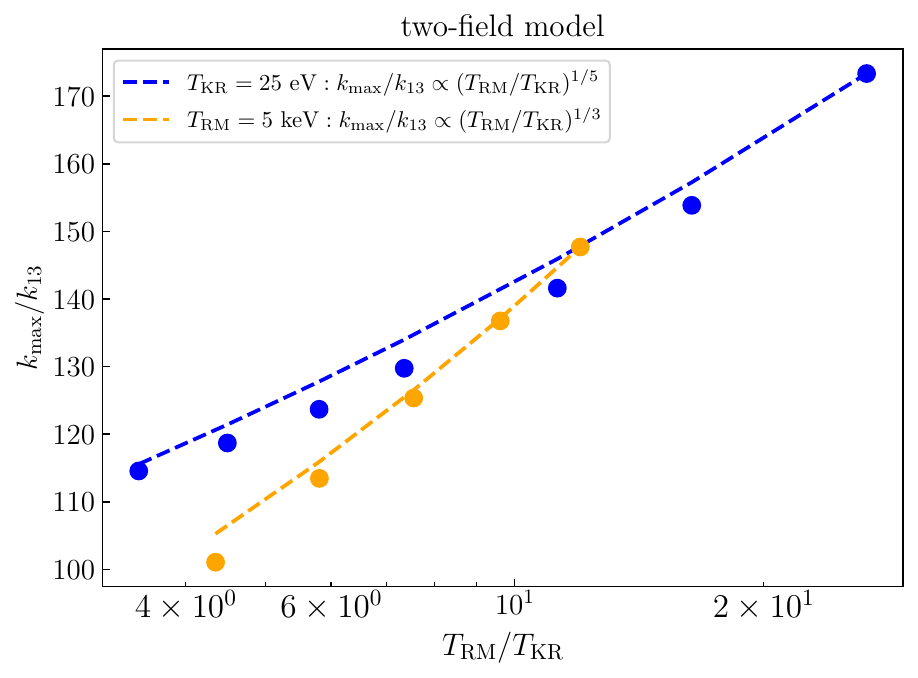}
     \includegraphics[width=0.495\linewidth]{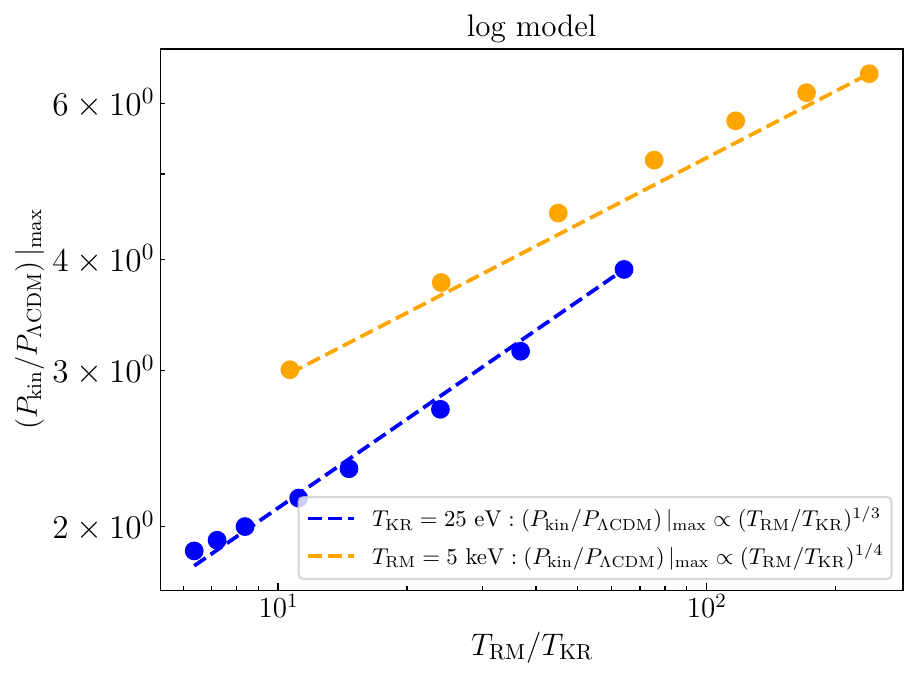}
    \includegraphics[width=0.495\linewidth]{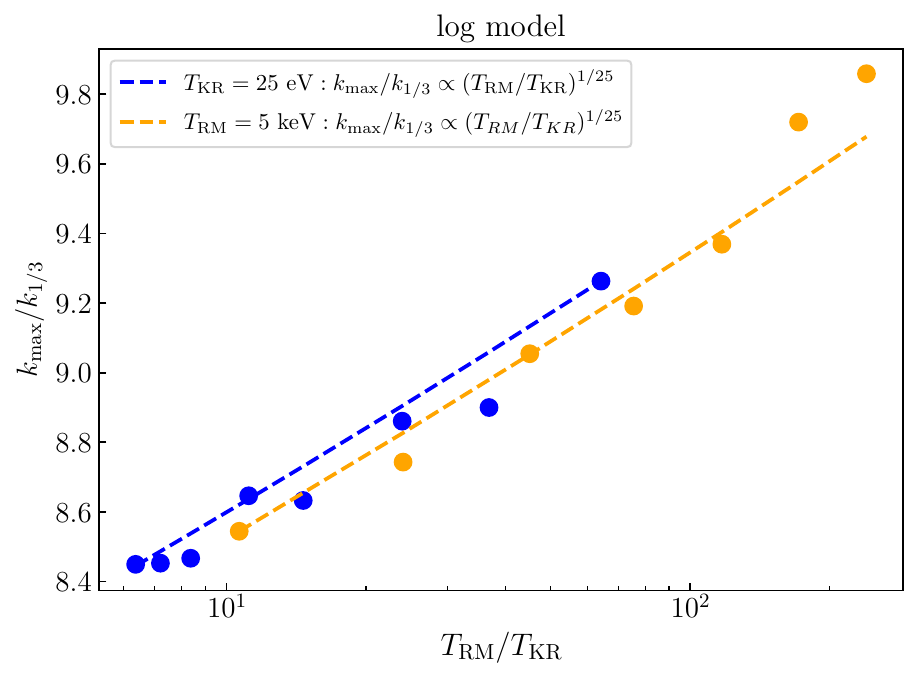}
\caption{
\textbf{Top Left:} The scaling of the maximum enhancement ${\rm max}( P_\mathrm{kin}/P_{\Lambda\mathrm{CDM}})$ as a function of the duration of axion kination domination $T_{\rm RM}/T_{\rm KR}$ for fixed $T_{\rm KR} = 25 \eV$ (blue) and fixed $T_{\rm RM} = 5 \keV$ (orange) in the two-field model. Dashed lines indicate the approximate scaling. \textbf{Top Right:} The scaling of the ratio $k_{\rm max}/k_{1/3}$  as a function of duration of axion kination domination in the two-field model, for fixed $T_{\rm KR} = 25 \eV$ (blue) and fixed $T_{\rm RM} = 5 \keV$ (orange). 
\textbf{Bottom Left:} Same as top left for the log model. 
\textbf{Bottom Right:} Same as top right for the log model.
}
\label{fig:MPSscaling}
\end{figure}

In the top left panel of Fig.~\ref{fig:MPSscaling}, we show the approximate scaling of the maximum enhancement of the MPS with respect to the standard $\Lambda {\rm CDM}$ cosmology, $(P_{\rm kin}/ P_{\Lambda {\rm CDM}})|_{\rm max}$, for the two-field model.  Orange points and lines show numerical values and power-law fits, respectively, for early-starting axion domination ($T_\mathrm{RM} = 5$ keV), while blue points and lines show the corresponding quantities for late-ending axion domination ($T_\mathrm{KR} = 25$ eV). The difference between these two curves thus highlights the dependence on the timing of axion domination.  
We see in Fig.~\ref{fig:MPSscaling} that the dependence of the maximum enhancement on the duration of modified expansion is nearly linear, $\propto (T_{\rm RM}/T_{\rm KR})^{7/6}$. This scaling of   $(P_{\rm kin}/ P_{\Lambda {\rm CDM}})|_{\rm max}$ with duration is insensitive to the timing of the feature, while the overall magnitude of the enhancement is mildly sensitive to the timing.

For the two-field model, we show the relationship between the comoving wave number $k_{\rm max}$ that receives the maximal enhancement and the mode $k_{1/3}$ that enters the horizon at $a_{1/3}$ in the top right panel, as a function of the duration of axion kination domination.  This ratio is weakly dependent on the duration of axion kination domination, and depends in detail on the timing.

The bottom panels show the same quantities for the log model. In contrast to the two-field model, the matter perturbations experience the bulk of their enhanced growth during the kination-dominated phase.  Since the total duration of the kination epoch is approximately $T_{1/3}/T_{\rm KR} \approx (T_{\rm RM}/T_{\rm KR})^{1/3}$, we see from the bottom right panel that the maximum growth does indeed roughly scale with the duration of the kination epoch. Meanwhile the ratio $k_{\rm max}/k_{1/3}$  is only very weakly dependent on the duration of axion kination domination.

\bibliographystyle{jhep}
\bibliography{kination}

\providecommand{\noopsort}[1]{}\providecommand{\singleletter}[1]{#1}%

\providecommand{\href}[2]{#2}\begingroup\raggedright\begin{thebibliography}{100}

\bibitem{Yeh:2022heq}
T.-H.~Yeh, J.~Shelton, K.A.~Olive and B.D.~Fields, \emph{{Probing physics beyond the standard model: limits from BBN and the CMB independently and combined}}, \href{https://doi.org/10.1088/1475-7516/2022/10/046}{\emph{JCAP} {\bfseries 10} (2022) 046} [\href{https://arxiv.org/abs/2207.13133}{{\ttfamily 2207.13133}}].

\bibitem{Sobotka:2022vrr}
A.C.~Sobotka, A.L.~Erickcek and T.L.~Smith, \emph{{Was entropy conserved between BBN and recombination?}}, \href{https://doi.org/10.1103/PhysRevD.107.023525}{\emph{Phys. Rev. D} {\bfseries 107} (2023) 023525} [\href{https://arxiv.org/abs/2207.14308}{{\ttfamily 2207.14308}}].

\bibitem{Sobotka:2023bzr}
A.C.~Sobotka, A.L.~Erickcek and T.L.~Smith, \emph{{Comprehensive Constraints on Dark Radiation Injection After BBN}},  \href{https://arxiv.org/abs/2312.13235}{{\ttfamily 2312.13235}}.

\bibitem{Co:2021lkc}
R.T.~Co, D.~Dunsky, N.~Fernandez, A.~Ghalsasi, L.J.~Hall, K.~Harigaya et~al., \emph{{Gravitational wave and CMB probes of axion kination}}, \href{https://doi.org/10.1007/JHEP09(2022)116}{\emph{JHEP} {\bfseries 09} (2022) 116} [\href{https://arxiv.org/abs/2108.09299}{{\ttfamily 2108.09299}}].

\bibitem{Sobotka:2024tat}
A.C.~Sobotka, A.L.~Erickcek and T.L.~Smith, \emph{{Signatures of very early dark energy in the matter power spectrum}}, \href{https://doi.org/10.1103/9bd9-fzwh}{\emph{Phys. Rev. D} {\bfseries 111} (2025) 123522} [\href{https://arxiv.org/abs/2409.06778}{{\ttfamily 2409.06778}}].

\bibitem{Co:2019wyp}
R.T.~Co and K.~Harigaya, \emph{{Axiogenesis}}, \href{https://doi.org/10.1103/PhysRevLett.124.111602}{\emph{Phys. Rev. Lett.} {\bfseries 124} (2020) 111602} [\href{https://arxiv.org/abs/1910.02080}{{\ttfamily 1910.02080}}].

\bibitem{Domcke:2020kcp}
V.~Domcke, Y.~Ema, K.~Mukaida and M.~Yamada, \emph{{Spontaneous Baryogenesis from Axions with Generic Couplings}}, \href{https://doi.org/10.1007/JHEP08(2020)096}{\emph{JHEP} {\bfseries 08} (2020) 096} [\href{https://arxiv.org/abs/2006.03148}{{\ttfamily 2006.03148}}].

\bibitem{Co:2020xlh}
R.T.~Co, L.J.~Hall and K.~Harigaya, \emph{{Predictions for Axion Couplings from ALP Cogenesis}}, \href{https://doi.org/10.1007/JHEP01(2021)172}{\emph{JHEP} {\bfseries 01} (2021) 172} [\href{https://arxiv.org/abs/2006.04809}{{\ttfamily 2006.04809}}].

\bibitem{Co:2020jtv}
R.T.~Co, N.~Fernandez, A.~Ghalsasi, L.J.~Hall and K.~Harigaya, \emph{{Lepto-Axiogenesis}}, \href{https://doi.org/10.1007/JHEP03(2021)017}{\emph{JHEP} {\bfseries 03} (2021) 017} [\href{https://arxiv.org/abs/2006.05687}{{\ttfamily 2006.05687}}].

\bibitem{Harigaya:2021txz}
K.~Harigaya and I.R.~Wang, \emph{{Axiogenesis from $SU(2)_R$ phase transition}}, \href{https://doi.org/10.1007/JHEP10(2021)022}{\emph{JHEP} {\bfseries 10} (2021) 022} [\href{https://arxiv.org/abs/2107.09679}{{\ttfamily 2107.09679}}].

\bibitem{Chakraborty:2021fkp}
S.~Chakraborty, T.H.~Jung and T.~Okui, \emph{{Composite neutrinos and the QCD axion: Baryogenesis, dark matter, small Dirac neutrino masses, and vanishing neutron electric dipole moment}}, \href{https://doi.org/10.1103/PhysRevD.105.015024}{\emph{Phys. Rev. D} {\bfseries 105} (2022) 015024} [\href{https://arxiv.org/abs/2108.04293}{{\ttfamily 2108.04293}}].

\bibitem{Kawamura:2021xpu}
J.~Kawamura and S.~Raby, \emph{{Lepto-axiogenesis in minimal SUSY KSVZ model}}, \href{https://doi.org/10.1007/JHEP04(2022)116}{\emph{JHEP} {\bfseries 04} (2022) 116} [\href{https://arxiv.org/abs/2109.08605}{{\ttfamily 2109.08605}}].

\bibitem{Co:2021qgl}
R.T.~Co, K.~Harigaya, Z.~Johnson and A.~Pierce, \emph{{R-parity violation axiogenesis}}, \href{https://doi.org/10.1007/JHEP11(2021)210}{\emph{JHEP} {\bfseries 11} (2021) 210} [\href{https://arxiv.org/abs/2110.05487}{{\ttfamily 2110.05487}}].

\bibitem{Co:2022aav}
R.T.~Co, T.~Gherghetta and K.~Harigaya, \emph{{Axiogenesis with a heavy QCD axion}}, \href{https://doi.org/10.1007/JHEP10(2022)121}{\emph{JHEP} {\bfseries 10} (2022) 121} [\href{https://arxiv.org/abs/2206.00678}{{\ttfamily 2206.00678}}].

\bibitem{Barnes:2022ren}
P.~Barnes, R.T.~Co, K.~Harigaya and A.~Pierce, \emph{{Lepto-axiogenesis and the scale of supersymmetry}}, \href{https://doi.org/10.1007/JHEP05(2023)114}{\emph{JHEP} {\bfseries 05} (2023) 114} [\href{https://arxiv.org/abs/2208.07878}{{\ttfamily 2208.07878}}].

\bibitem{Co:2022kul}
R.T.~Co, V.~Domcke and K.~Harigaya, \emph{{Baryogenesis from decaying magnetic helicity in axiogenesis}}, \href{https://doi.org/10.1007/JHEP07(2023)179}{\emph{JHEP} {\bfseries 07} (2023) 179} [\href{https://arxiv.org/abs/2211.12517}{{\ttfamily 2211.12517}}].

\bibitem{Berbig:2023uzs}
M.~Berbig, \emph{{Diraxiogenesis}}, \href{https://doi.org/10.1007/JHEP01(2024)061}{\emph{JHEP} {\bfseries 01} (2024) 061} [\href{https://arxiv.org/abs/2307.14121}{{\ttfamily 2307.14121}}].

\bibitem{Chun:2023eqc}
E.J.~Chun and T.H.~Jung, \emph{{Leptogenesis driven by a Majoron}}, \href{https://doi.org/10.1103/PhysRevD.109.095004}{\emph{Phys. Rev. D} {\bfseries 109} (2024) 095004} [\href{https://arxiv.org/abs/2311.09005}{{\ttfamily 2311.09005}}].

\bibitem{Barnes:2024jap}
P.~Barnes, R.T.~Co, K.~Harigaya and A.~Pierce, \emph{{Lepto-axiogenesis with light right-handed neutrinos}}, \href{https://doi.org/10.1007/JHEP08(2025)004}{\emph{JHEP} {\bfseries 08} (2025) 004} [\href{https://arxiv.org/abs/2402.10263}{{\ttfamily 2402.10263}}].

\bibitem{Wada:2024cbe}
J.~Wada, \emph{{Majoron-driven leptogenesis in gauged U(1)L\ensuremath{\mu}-L\ensuremath{\tau} model}}, \href{https://doi.org/10.1103/PhysRevD.110.103510}{\emph{Phys. Rev. D} {\bfseries 110} (2024) 103510} [\href{https://arxiv.org/abs/2404.10283}{{\ttfamily 2404.10283}}].

\bibitem{Datta:2024xhg}
A.~Datta, S.K.~Manna and A.~Sil, \emph{{Spontaneous leptogenesis with sub-GeV axionlike particles}}, \href{https://doi.org/10.1103/PhysRevD.110.095035}{\emph{Phys. Rev. D} {\bfseries 110} (2024) 095035} [\href{https://arxiv.org/abs/2405.07003}{{\ttfamily 2405.07003}}].

\bibitem{Co:2019jts}
R.T.~Co, L.J.~Hall and K.~Harigaya, \emph{{Axion Kinetic Misalignment Mechanism}}, \href{https://doi.org/10.1103/PhysRevLett.124.251802}{\emph{Phys. Rev. Lett.} {\bfseries 124} (2020) 251802} [\href{https://arxiv.org/abs/1910.14152}{{\ttfamily 1910.14152}}].

\bibitem{Co:2020dya}
R.T.~Co, L.J.~Hall, K.~Harigaya, K.A.~Olive and S.~Verner, \emph{{Axion Kinetic Misalignment and Parametric Resonance from Inflation}}, \href{https://doi.org/10.1088/1475-7516/2020/08/036}{\emph{JCAP} {\bfseries 08} (2020) 036} [\href{https://arxiv.org/abs/2004.00629}{{\ttfamily 2004.00629}}].

\bibitem{Eroncel:2022vjg}
C.~Er{\"o}ncel, R.~Sato, G.~Servant and P.~S{\o}rensen, \emph{{ALP dark matter from kinetic fragmentation: opening up the parameter window}}, \href{https://doi.org/10.1088/1475-7516/2022/10/053}{\emph{JCAP} {\bfseries 10} (2022) 053} [\href{https://arxiv.org/abs/2206.14259}{{\ttfamily 2206.14259}}].

\bibitem{Eroncel:2025qlk}
C.~Er{\"o}ncel, Y.~Gouttenoire, R.~Sato, G.~Servant and P.~Simakachorn, \emph{{A New Source for (QCD) Axion Dark Matter Production: Curvature-Induced}},  \href{https://arxiv.org/abs/2503.04880}{{\ttfamily 2503.04880}}.

\bibitem{Bodas:2025eca}
A.~Bodas, R.T.~Co, A.~Ghalsasi, K.~Harigaya and L.-T.~Wang, \emph{{Acoustic misalignment mechanism for axion dark matter}}, \href{https://doi.org/10.1007/JHEP08(2025)131}{\emph{JHEP} {\bfseries 08} (2025) 131} [\href{https://arxiv.org/abs/2503.04888}{{\ttfamily 2503.04888}}].

\bibitem{Co:2021rhi}
R.T.~Co, K.~Harigaya and A.~Pierce, \emph{{Gravitational waves and dark photon dark matter from axion rotations}}, \href{https://doi.org/10.1007/JHEP12(2021)099}{\emph{JHEP} {\bfseries 12} (2021) 099} [\href{https://arxiv.org/abs/2104.02077}{{\ttfamily 2104.02077}}].

\bibitem{Madge:2021abk}
E.~Madge, W.~Ratzinger, D.~Schmitt and P.~Schwaller, \emph{{Audible axions with a booster: Stochastic gravitational waves from rotating ALPs}}, \href{https://doi.org/10.21468/SciPostPhys.12.5.171}{\emph{SciPost Phys.} {\bfseries 12} (2022) 171} [\href{https://arxiv.org/abs/2111.12730}{{\ttfamily 2111.12730}}].

\bibitem{Gouttenoire:2021wzu}
Y.~Gouttenoire, G.~Servant and P.~Simakachorn, \emph{{Revealing the Primordial Irreducible Inflationary Gravitational-Wave Background with a Spinning Peccei-Quinn Axion}},  \href{https://arxiv.org/abs/2108.10328}{{\ttfamily 2108.10328}}.

\bibitem{Harigaya:2023mhl}
K.~Harigaya, K.~Inomata and T.~Terada, \emph{{Gravitational wave production from axion rotations right after a transition to kination}}, \href{https://doi.org/10.1103/PhysRevD.108.L081303}{\emph{Phys. Rev. D} {\bfseries 108} (2023) L081303} [\href{https://arxiv.org/abs/2305.14242}{{\ttfamily 2305.14242}}].

\bibitem{Bodas:2025wef}
A.~Bodas, K.~Harigaya, K.~Inomata, T.~Terada and L.-T.~Wang, \emph{{Anisotropic Gravitational Waves from Anisotropic Axion Rotation}},  \href{https://arxiv.org/abs/2508.08249}{{\ttfamily 2508.08249}}.

\bibitem{Co:2024oek}
R.T.~Co, N.~Fernandez, A.~Ghalsasi, K.~Harigaya and J.~Shelton, \emph{{Axion baryogenesis puts a new spin on the Hubble tension}}, \href{https://doi.org/10.1103/PhysRevD.110.083534}{\emph{Phys. Rev. D} {\bfseries 110} (2024) 083534} [\href{https://arxiv.org/abs/2405.12268}{{\ttfamily 2405.12268}}].

\bibitem{Bechtol:2022koa}
K.~Bechtol et~al., \emph{{Snowmass2021 Cosmic Frontier White Paper: Dark Matter Physics from Halo Measurements}},  in \emph{{Snowmass 2021}}, 3, 2022 [\href{https://arxiv.org/abs/2203.07354}{{\ttfamily 2203.07354}}].

\bibitem{Mao:2022fyx}
Y.-Y.~Mao et~al., \emph{{Snowmass2021: Vera C. Rubin Observatory as a Flagship Dark Matter Experiment}},  \href{https://arxiv.org/abs/2203.07252}{{\ttfamily 2203.07252}}.

\bibitem{2022ApJ...940L..14N}
R.P.~{Naidu}, P.A.~{Oesch}, P.~{van Dokkum}, E.J.~{Nelson}, K.A.~{Suess}, G.~{Brammer} et~al., \emph{{Two Remarkably Luminous Galaxy Candidates at z {\ensuremath{\approx}} 10-12 Revealed by JWST}}, \href{https://doi.org/10.3847/2041-8213/ac9b22}{\emph{Ap.J.L.} {\bfseries 940} (2022) L14} [\href{https://arxiv.org/abs/2207.09434}{{\ttfamily 2207.09434}}].

\bibitem{2022ApJ...938L..15C}
M.~{Castellano}, A.~{Fontana}, T.~{Treu}, P.~{Santini}, E.~{Merlin}, N.~{Leethochawalit} et~al., \emph{{Early Results from GLASS-JWST. III. Galaxy Candidates at z 9-15}}, \href{https://doi.org/10.3847/2041-8213/ac94d0}{\emph{Ap.J.L..} {\bfseries 938} (2022) L15} [\href{https://arxiv.org/abs/2207.09436}{{\ttfamily 2207.09436}}].

\bibitem{2024Natur.633..318C}
S.~{Carniani}, K.~{Hainline}, F.~{D'Eugenio}, D.J.~{Eisenstein}, P.~{Jakobsen}, J.~{Witstok} et~al., \emph{{Spectroscopic confirmation of two luminous galaxies at a redshift of 14}}, \href{https://doi.org/10.1038/s41586-024-07860-9}{\emph{Nature} {\bfseries 633} (2024) 318} [\href{https://arxiv.org/abs/2405.18485}{{\ttfamily 2405.18485}}].

\bibitem{2023MNRAS.519.1201A}
H.~{Atek}, M.~{Shuntov}, L.J.~{Furtak}, J.~{Richard}, J.-P.~{Kneib}, G.~{Mahler} et~al., \emph{{Revealing galaxy candidates out to z 16 with JWST observations of the lensing cluster SMACS0723}}, \href{https://doi.org/10.1093/mnras/stac3144}{\emph{MNRAS} {\bfseries 519} (2023) 1201} [\href{https://arxiv.org/abs/2207.12338}{{\ttfamily 2207.12338}}].

\bibitem{2023Natur.616..266L}
I.~{Labb{\'e}}, P.~{van Dokkum}, E.~{Nelson}, R.~{Bezanson}, K.A.~{Suess}, J.~{Leja} et~al., \emph{{A population of red candidate massive galaxies 600 Myr after the Big Bang}}, \href{https://doi.org/10.1038/s41586-023-05786-2}{\emph{Nature} {\bfseries 616} (2023) 266} [\href{https://arxiv.org/abs/2207.12446}{{\ttfamily 2207.12446}}].

\bibitem{2025arXiv250821748J}
I.~{Juod{\v{z}}balis}, C.~{Marconcini}, F.~{D'Eugenio}, R.~{Maiolino}, A.~{Marconi}, H.~{{\"U}bler} et~al., \emph{{A direct black hole mass measurement in a Little Red Dot at the Epoch of Reionization}}, \href{https://doi.org/10.48550/arXiv.2508.21748}{\emph{arXiv e-prints} (2025) arXiv:2508.21748} [\href{https://arxiv.org/abs/2508.21748}{{\ttfamily 2508.21748}}].

\bibitem{2024NatAs...8..126B}
{\'A}.~{Bogd{\'a}n}, A.D.~{Goulding}, P.~{Natarajan}, O.E.~{Kov{\'a}cs}, G.R.~{Tremblay}, U.~{Chadayammuri} et~al., \emph{{Evidence for heavy-seed origin of early supermassive black holes from a z {\ensuremath{\approx}} 10 X-ray quasar}}, \href{https://doi.org/10.1038/s41550-023-02111-9}{\emph{Nature Astronomy} {\bfseries 8} (2024) 126} [\href{https://arxiv.org/abs/2305.15458}{{\ttfamily 2305.15458}}].

\bibitem{2023ApJ...957L...7K}
V.~{Kokorev}, S.~{Fujimoto}, I.~{Labbe}, J.E.~{Greene}, R.~{Bezanson}, P.~{Dayal} et~al., \emph{{UNCOVER: A NIRSpec Identification of a Broad-line AGN at z = 8.50}}, \href{https://doi.org/10.3847/2041-8213/ad037a}{\emph{ApJL} {\bfseries 957} (2023) L7} [\href{https://arxiv.org/abs/2308.11610}{{\ttfamily 2308.11610}}].

\bibitem{2023ApJ...953L..29L}
R.L.~{Larson}, S.L.~{Finkelstein}, D.D.~{Kocevski}, T.A.~{Hutchison}, J.R.~{Trump}, P.~{Arrabal Haro} et~al., \emph{{A CEERS Discovery of an Accreting Supermassive Black Hole 570 Myr after the Big Bang: Identifying a Progenitor of Massive z > 6 Quasars}}, \href{https://doi.org/10.3847/2041-8213/ace619}{\emph{ApJL} {\bfseries 953} (2023) L29} [\href{https://arxiv.org/abs/2303.08918}{{\ttfamily 2303.08918}}].

\bibitem{2024Natur.628...57F}
L.J.~{Furtak}, I.~{Labb{\'e}}, A.~{Zitrin}, J.E.~{Greene}, P.~{Dayal}, I.~{Chemerynska} et~al., \emph{{A high black-hole-to-host mass ratio in a lensed AGN in the early Universe}}, \href{https://doi.org/10.1038/s41586-024-07184-8}{\emph{Nature} {\bfseries 628} (2024) 57} [\href{https://arxiv.org/abs/2308.05735}{{\ttfamily 2308.05735}}].

\bibitem{2023MNRAS.525.1353J}
I.~{Juod{\v{z}}balis}, C.J.~{Conselice}, M.~{Singh}, N.~{Adams}, K.~{Ormerod}, T.~{Harvey} et~al., \emph{{EPOCHS VII: discovery of high-redshift (6.5 < z < 12) AGN candidates in JWST ERO and PEARLS data}}, \href{https://doi.org/10.1093/mnras/stad2396}{\emph{MNRAS} {\bfseries 525} (2023) 1353} [\href{https://arxiv.org/abs/2307.07535}{{\ttfamily 2307.07535}}].

\bibitem{2024arXiv241018763N}
L.~{Napolitano}, M.~{Castellano}, L.~{Pentericci}, C.~{Vignali}, R.~{Gilli}, A.~{Fontana} et~al., \emph{{The dual nature of GHZ9: coexisting AGN and star formation activity in a remote X-ray source at z=10.145}}, \href{https://doi.org/10.48550/arXiv.2410.18763}{\emph{arXiv e-prints} (2024) arXiv:2410.18763} [\href{https://arxiv.org/abs/2410.18763}{{\ttfamily 2410.18763}}].

\bibitem{Qin:2025ymc}
W.~Qin, S.~Kumar, P.~Natarajan and N.~Weiner, \emph{{Not-quite-primordial black holes}},  \href{https://arxiv.org/abs/2506.13858}{{\ttfamily 2506.13858}}.

\bibitem{Affleck:1984fy}
I.~Affleck and M.~Dine, \emph{{A New Mechanism for Baryogenesis}}, \href{https://doi.org/10.1016/0550-3213(85)90021-5}{\emph{Nucl. Phys. B} {\bfseries 249} (1985) 361}.

\bibitem{Dine:1995kz}
M.~Dine, L.~Randall and S.D.~Thomas, \emph{{Baryogenesis from flat directions of the supersymmetric standard model}}, \href{https://doi.org/10.1016/0550-3213(95)00538-2}{\emph{Nucl. Phys. B} {\bfseries 458} (1996) 291} [\href{https://arxiv.org/abs/hep-ph/9507453}{{\ttfamily hep-ph/9507453}}].

\bibitem{Domcke:2022wpb}
V.~Domcke, K.~Harigaya and K.~Mukaida, \emph{{Charge transfer between rotating complex scalar fields}}, \href{https://doi.org/10.1007/JHEP08(2022)234}{\emph{JHEP} {\bfseries 08} (2022) 234} [\href{https://arxiv.org/abs/2205.00942}{{\ttfamily 2205.00942}}].

\bibitem{Moxhay:1984am}
P.~Moxhay and K.~Yamamoto, \emph{{{Peccei-Quinn} Symmetry Breaking by Radiative Corrections in Supergravity}}, \href{https://doi.org/10.1016/0370-2693(85)91655-7}{\emph{Phys. Lett. B} {\bfseries 151} (1985) 363}.

\bibitem{Ma:1995ey}
C.-P.~Ma and E.~Bertschinger, \emph{{Cosmological perturbation theory in the synchronous and conformal Newtonian gauges}}, \href{https://doi.org/10.1086/176550}{\emph{Astrophys. J.} {\bfseries 455} (1995) 7} [\href{https://arxiv.org/abs/astro-ph/9506072}{{\ttfamily astro-ph/9506072}}].

\bibitem{Blas:2011rf}
D.~Blas, J.~Lesgourgues and T.~Tram, \emph{{The Cosmic Linear Anisotropy Solving System (CLASS) II: Approximation schemes}}, \href{https://doi.org/10.1088/1475-7516/2011/07/034}{\emph{JCAP} {\bfseries 07} (2011) 034} [\href{https://arxiv.org/abs/1104.2933}{{\ttfamily 1104.2933}}].

\bibitem{Redmond:2018xty}
K.~Redmond, A.~Trezza and A.L.~Erickcek, \emph{{Growth of Dark Matter Perturbations during Kination}}, \href{https://doi.org/10.1103/PhysRevD.98.063504}{\emph{Phys. Rev. D} {\bfseries 98} (2018) 063504} [\href{https://arxiv.org/abs/1807.01327}{{\ttfamily 1807.01327}}].

\bibitem{Eroncel:2025bcb}
C.~Er{\"o}ncel, Y.~Gouttenoire, R.~Sato, G.~Servant and P.~Simakachorn, \emph{{Universal Bound on the Duration of a Kination Era}}, \href{https://doi.org/10.1103/k7ty-gwjg}{\emph{Phys. Rev. Lett.} {\bfseries 135} (2025) 101002} [\href{https://arxiv.org/abs/2501.17226}{{\ttfamily 2501.17226}}].

\bibitem{Burns:2023sgx}
A.-K.~Burns, T.M.P.~Tait and M.~Valli, \emph{{PRyMordial: the first three minutes, within and beyond the standard model}}, \href{https://doi.org/10.1140/epjc/s10052-024-12442-0}{\emph{Eur. Phys. J. C} {\bfseries 84} (2024) 86} [\href{https://arxiv.org/abs/2307.07061}{{\ttfamily 2307.07061}}].

\bibitem{Planck:2018vyg}
{\scshape Planck} collaboration, \emph{{Planck 2018 results. VI. Cosmological parameters}}, \href{https://doi.org/10.1051/0004-6361/201833910}{\emph{Astron. Astrophys.} {\bfseries 641} (2020) A6} [\href{https://arxiv.org/abs/1807.06209}{{\ttfamily 1807.06209}}].

\bibitem{2020ApJ...896...77H}
T.~{Hsyu}, R.J.~{Cooke}, J.X.~{Prochaska} and M.~{Bolte}, \emph{{The PHLEK Survey: A New Determination of the Primordial Helium Abundance}}, \href{https://doi.org/10.3847/1538-4357/ab91af}{\emph{Ap.J.} {\bfseries 896} (2020) 77} [\href{https://arxiv.org/abs/2005.12290}{{\ttfamily 2005.12290}}].

\bibitem{Aver:2020fon}
E.~Aver, D.A.~Berg, K.A.~Olive, R.W.~Pogge, J.J.~Salzer and E.D.~Skillman, \emph{{Improving helium abundance determinations with Leo P as a case study}}, \href{https://doi.org/10.1088/1475-7516/2021/03/027}{\emph{JCAP} {\bfseries 03} (2021) 027} [\href{https://arxiv.org/abs/2010.04180}{{\ttfamily 2010.04180}}].

\bibitem{Riemer-Sorensen:2017vxj}
S.~Riemer-S{\o}rensen and E.S.~Jenssen, \emph{{Nucleosynthesis Predictions and High-Precision Deuterium Measurements}}, \href{https://doi.org/10.3390/universe3020044}{\emph{Universe} {\bfseries 3} (2017) 44} [\href{https://arxiv.org/abs/1705.03653}{{\ttfamily 1705.03653}}].

\bibitem{Cooke:2017cwo}
R.J.~Cooke, M.~Pettini and C.C.~Steidel, \emph{{One Percent Determination of the Primordial Deuterium Abundance}}, \href{https://doi.org/10.3847/1538-4357/aaab53}{\emph{Astrophys. J.} {\bfseries 855} (2018) 102} [\href{https://arxiv.org/abs/1710.11129}{{\ttfamily 1710.11129}}].

\bibitem{Pitrou:2020etk}
C.~Pitrou, A.~Coc, J.-P.~Uzan and E.~Vangioni, \emph{{A new tension in the cosmological model from primordial deuterium?}}, \href{https://doi.org/10.1093/mnras/stab135}{\emph{Mon. Not. Roy. Astron. Soc.} {\bfseries 502} (2021) 2474} [\href{https://arxiv.org/abs/2011.11320}{{\ttfamily 2011.11320}}].

\bibitem{Brinckmann:2018cvx}
T.~Brinckmann and J.~Lesgourgues, \emph{{MontePython 3: boosted MCMC sampler and other features}}, \href{https://doi.org/10.1016/j.dark.2018.100260}{\emph{Phys. Dark Univ.} {\bfseries 24} (2019) 100260} [\href{https://arxiv.org/abs/1804.07261}{{\ttfamily 1804.07261}}].

\bibitem{Beutler:2011hx}
F.~Beutler, C.~Blake, M.~Colless, D.H.~Jones, L.~Staveley-Smith, L.~Campbell et~al., \emph{{The 6dF Galaxy Survey: Baryon Acoustic Oscillations and the Local Hubble Constant}}, \href{https://doi.org/10.1111/j.1365-2966.2011.19250.x}{\emph{Mon. Not. Roy. Astron. Soc.} {\bfseries 416} (2011) 3017} [\href{https://arxiv.org/abs/1106.3366}{{\ttfamily 1106.3366}}].

\bibitem{Ross:2014qpa}
A.J.~Ross, L.~Samushia, C.~Howlett, W.J.~Percival, A.~Burden and M.~Manera, \emph{{The clustering of the SDSS DR7 main Galaxy sample \textendash{} I. A 4 per cent distance measure at $z = 0.15$}}, \href{https://doi.org/10.1093/mnras/stv154}{\emph{Mon. Not. Roy. Astron. Soc.} {\bfseries 449} (2015) 835} [\href{https://arxiv.org/abs/1409.3242}{{\ttfamily 1409.3242}}].

\bibitem{BOSS:2016wmc}
{\scshape BOSS} collaboration, \emph{{The clustering of galaxies in the completed SDSS-III Baryon Oscillation Spectroscopic Survey: cosmological analysis of the DR12 galaxy sample}}, \href{https://doi.org/10.1093/mnras/stx721}{\emph{Mon. Not. Roy. Astron. Soc.} {\bfseries 470} (2017) 2617} [\href{https://arxiv.org/abs/1607.03155}{{\ttfamily 1607.03155}}].

\bibitem{Gelman:1992zz}
A.~Gelman and D.B.~Rubin, \emph{{Inference from Iterative Simulation Using Multiple Sequences}}, \href{https://doi.org/10.1214/ss/1177011136}{\emph{Statist. Sci.} {\bfseries 7} (1992) 457}.

\bibitem{Bird:2010mp}
S.~Bird, H.V.~Peiris, M.~Viel and L.~Verde, \emph{{Minimally Parametric Power Spectrum Reconstruction from the Lyman-alpha Forest}}, \href{https://doi.org/10.1111/j.1365-2966.2011.18245.x}{\emph{Mon. Not. Roy. Astron. Soc.} {\bfseries 413} (2011) 1717} [\href{https://arxiv.org/abs/1010.1519}{{\ttfamily 1010.1519}}].

\bibitem{Fernandez:2023grg}
M.A.~Fernandez, S.~Bird and M.-F.~Ho, \emph{{Cosmological constraints from the eBOSS Lyman-\ensuremath{\alpha} forest using the PRIYA simulations}}, \href{https://doi.org/10.1088/1475-7516/2024/07/029}{\emph{JCAP} {\bfseries 07} (2024) 029} [\href{https://arxiv.org/abs/2309.03943}{{\ttfamily 2309.03943}}].

\bibitem{eBOSS:2018qyj}
{\scshape eBOSS} collaboration, \emph{{The one-dimensional power spectrum from the SDSS DR14 Ly$\alpha$ forests}}, \href{https://doi.org/10.1088/1475-7516/2019/07/017}{\emph{JCAP} {\bfseries 07} (2019) 017} [\href{https://arxiv.org/abs/1812.03554}{{\ttfamily 1812.03554}}].

\bibitem{Dekker:2024nkb}
A.~Dekker and A.~Kravtsov, \emph{{Constraints on blue and red tilted primordial power spectra using dwarf galaxy properties}}, \href{https://doi.org/10.1103/PhysRevD.111.063516}{\emph{Phys. Rev. D} {\bfseries 111} (2025) 063516} [\href{https://arxiv.org/abs/2407.04198}{{\ttfamily 2407.04198}}].

\bibitem{Graham:2024hah}
P.W.~Graham and H.~Ramani, \emph{{Constraints on dark matter from dynamical heating of stars in ultrafaint dwarfs. II. Substructure and the primordial power spectrum}}, \href{https://doi.org/10.1103/PhysRevD.110.075012}{\emph{Phys. Rev. D} {\bfseries 110} (2024) 075012} [\href{https://arxiv.org/abs/2404.01378}{{\ttfamily 2404.01378}}].

\bibitem{Esteban:2023xpk}
I.~Esteban, A.H.G.~Peter and S.Y.~Kim, \emph{{Milky~Way satellite velocities reveal the dark matter power spectrum at small scales}}, \href{https://doi.org/10.1103/PhysRevD.110.123013}{\emph{Phys. Rev. D} {\bfseries 110} (2024) 123013} [\href{https://arxiv.org/abs/2306.04674}{{\ttfamily 2306.04674}}].

\bibitem{Nadler:2025crd}
E.O.~Nadler, V.~Gluscevic and A.~Benson, \emph{{The Effects of Linear Matter Power Spectrum Enhancement on Dark Matter Substructure}},  \href{https://arxiv.org/abs/2507.16889}{{\ttfamily 2507.16889}}.

\bibitem{Gilman:2021gkj}
D.~Gilman, A.~Benson, J.~Bovy, S.~Birrer, T.~Treu and A.~Nierenberg, \emph{{The primordial matter power spectrum on sub-galactic scales}}, \href{https://doi.org/10.1093/mnras/stac670}{\emph{Mon. Not. Roy. Astron. Soc.} {\bfseries 512} (2022) 3163} [\href{https://arxiv.org/abs/2112.03293}{{\ttfamily 2112.03293}}].

\bibitem{Sabti:2021unj}
N.~Sabti, J.B.~Mu{\~n}oz and D.~Blas, \emph{{New Roads to the Small-scale Universe: Measurements of the Clustering of Matter with the High-redshift UV Galaxy Luminosity Function}}, \href{https://doi.org/10.3847/2041-8213/ac5e9c}{\emph{Astrophys. J. Lett.} {\bfseries 928} (2022) L20} [\href{https://arxiv.org/abs/2110.13161}{{\ttfamily 2110.13161}}].

\bibitem{Sabti:2023xwo}
N.~Sabti, J.B.~Mu{\~n}oz and M.~Kamionkowski, \emph{{Insights from HST into Ultramassive Galaxies and Early-Universe Cosmology}}, \href{https://doi.org/10.1103/PhysRevLett.132.061002}{\emph{Phys. Rev. Lett.} {\bfseries 132} (2024) 061002} [\href{https://arxiv.org/abs/2305.07049}{{\ttfamily 2305.07049}}].

\bibitem{Chluba:2015bqa}
J.~Chluba, J.~Hamann and S.P.~Patil, \emph{{Features and New Physical Scales in Primordial Observables: Theory and Observation}}, \href{https://doi.org/10.1142/S0218271815300232}{\emph{Int. J. Mod. Phys. D} {\bfseries 24} (2015) 1530023} [\href{https://arxiv.org/abs/1505.01834}{{\ttfamily 1505.01834}}].

\bibitem{Kogut:2011xw}
A.~Kogut et~al., \emph{{The Primordial Inflation Explorer (PIXIE): A Nulling Polarimeter for Cosmic Microwave Background Observations}}, \href{https://doi.org/10.1088/1475-7516/2011/07/025}{\emph{JCAP} {\bfseries 07} (2011) 025} [\href{https://arxiv.org/abs/1105.2044}{{\ttfamily 1105.2044}}].

\bibitem{Linde:1984ti}
A.D.~Linde, \emph{{GENERATION OF ISOTHERMAL DENSITY PERTURBATIONS IN THE INFLATIONARY UNIVERSE}}, {\emph{JETP Lett.} {\bfseries 40} (1984) 1333}.

\bibitem{Chung:2004nh}
D.J.H.~Chung, E.W.~Kolb, A.~Riotto and L.~Senatore, \emph{{Isocurvature constraints on gravitationally produced superheavy dark matter}}, \href{https://doi.org/10.1103/PhysRevD.72.023511}{\emph{Phys. Rev. D} {\bfseries 72} (2005) 023511} [\href{https://arxiv.org/abs/astro-ph/0411468}{{\ttfamily astro-ph/0411468}}].

\bibitem{Chung:2013rda}
D.J.H.~Chung, H.~Yoo and P.~Zhou, \emph{{Fermionic Isocurvature Perturbations}}, \href{https://doi.org/10.1103/PhysRevD.91.043516}{\emph{Phys. Rev. D} {\bfseries 91} (2015) 043516} [\href{https://arxiv.org/abs/1306.1966}{{\ttfamily 1306.1966}}].

\bibitem{Graham:2015rva}
P.W.~Graham, J.~Mardon and S.~Rajendran, \emph{{Vector Dark Matter from Inflationary Fluctuations}}, \href{https://doi.org/10.1103/PhysRevD.93.103520}{\emph{Phys. Rev. D} {\bfseries 93} (2016) 103520} [\href{https://arxiv.org/abs/1504.02102}{{\ttfamily 1504.02102}}].

\bibitem{Alonso-Alvarez:2018tus}
G.~Alonso-{\'A}lvarez and J.~Jaeckel, \emph{{Lightish but clumpy: scalar dark matter from inflationary fluctuations}}, \href{https://doi.org/10.1088/1475-7516/2018/10/022}{\emph{JCAP} {\bfseries 10} (2018) 022} [\href{https://arxiv.org/abs/1807.09785}{{\ttfamily 1807.09785}}].

\bibitem{Kolb:2023ydq}
E.W.~Kolb and A.J.~Long, \emph{{Cosmological gravitational particle production and its implications for cosmological relics}}, \href{https://doi.org/10.1103/RevModPhys.96.045005}{\emph{Rev. Mod. Phys.} {\bfseries 96} (2024) 045005} [\href{https://arxiv.org/abs/2312.09042}{{\ttfamily 2312.09042}}].

\bibitem{Chluba:2013dna}
J.~Chluba and D.~Grin, \emph{{CMB spectral distortions from small-scale isocurvature fluctuations}}, \href{https://doi.org/10.1093/mnras/stt1129}{\emph{Mon. Not. Roy. Astron. Soc.} {\bfseries 434} (2013) 1619} [\href{https://arxiv.org/abs/1304.4596}{{\ttfamily 1304.4596}}].

\bibitem{Buckley:2025zgh}
M.R.~Buckley, P.~Du, N.~Fernandez and M.J.~Weikert, \emph{{General Constraints on Isocurvature from the CMB and Ly-$\alpha$ Forest}},  \href{https://arxiv.org/abs/2502.20434}{{\ttfamily 2502.20434}}.

\bibitem{Rogers:2023upm}
K.K.~Rogers and V.~Poulin, \emph{{5{\ensuremath{\sigma}} tension between Planck cosmic microwave background and eBOSS Lyman-alpha forest and constraints on physics beyond {\ensuremath{\Lambda}}CDM}}, \href{https://doi.org/10.1103/PhysRevResearch.7.L012018}{\emph{Phys. Rev. Res.} {\bfseries 7} (2025) L012018} [\href{https://arxiv.org/abs/2311.16377}{{\ttfamily 2311.16377}}].

\bibitem{2023ApJ...955L..35S}
G.~{Sun}, C.-A.~{Faucher-Gigu{\`e}re}, C.C.~{Hayward}, X.~{Shen}, A.~{Wetzel} and R.K.~{Cochrane}, \emph{{Bursty Star Formation Naturally Explains the Abundance of Bright Galaxies at Cosmic Dawn}}, \href{https://doi.org/10.3847/2041-8213/acf85a}{\emph{ApJL} {\bfseries 955} (2023) L35} [\href{https://arxiv.org/abs/2307.15305}{{\ttfamily 2307.15305}}].

\bibitem{Feldmann:2024kwz}
R.~Feldmann et~al., \emph{{Elevated UV luminosity density at Cosmic Dawn explained by non-evolving, weakly mass-dependent star formation efficiency}}, \href{https://doi.org/10.1093/mnras/stae2633}{\emph{Mon. Not. Roy. Astron. Soc.} {\bfseries 536} (2024) 988} [\href{https://arxiv.org/abs/2407.02674}{{\ttfamily 2407.02674}}].

\bibitem{Ellis:2025xju}
J.~Ellis, M.~Fairbairn, J.~Urrutia and V.~Vaskonen, \emph{{Starlight from JWST: Implications for star formation and dark matter models}},  \href{https://arxiv.org/abs/2504.20043}{{\ttfamily 2504.20043}}.

\bibitem{Hezaveh:2016ltk}
Y.D.~Hezaveh et~al., \emph{{Detection of lensing substructure using ALMA observations of the dusty galaxy SDP.81}}, \href{https://doi.org/10.3847/0004-637X/823/1/37}{\emph{Astrophys. J.} {\bfseries 823} (2016) 37} [\href{https://arxiv.org/abs/1601.01388}{{\ttfamily 1601.01388}}].

\bibitem{Despali:2024ihn}
G.~Despali, F.M.~Heinze, C.D.~Fassnacht, S.~Vegetti, C.~Spingola, R.~Klessen et~al., \emph{{Detecting low-mass haloes with strong gravitational lensing - II. Constraints on the density profiles of two detected subhaloes}}, \href{https://doi.org/10.1051/0004-6361/202451546}{\emph{Astron. Astrophys.} {\bfseries 699} (2025) A222} [\href{https://arxiv.org/abs/2407.12910}{{\ttfamily 2407.12910}}].

\bibitem{Enzi:2024ygw}
W.J.R.~Enzi, C.M.~Krawczyk, D.J.~Ballard and T.E.~Collett, \emph{{The overconcentrated dark halo in the strong lens SDSS J0946+1006 is a subhalo: evidence for self interacting dark matter?}},  \href{https://arxiv.org/abs/2411.08565}{{\ttfamily 2411.08565}}.

\bibitem{Sengul:2021lxe}
A.{\c{C}}.~{\c{S}}eng{\"u}l, C.~Dvorkin, B.~Ostdiek and A.~Tsang, \emph{{Substructure detection reanalysed: dark perturber shown to be a line-of-sight halo}}, \href{https://doi.org/10.1093/mnras/stac1967}{\emph{Mon. Not. Roy. Astron. Soc.} {\bfseries 515} (2022) 4391} [\href{https://arxiv.org/abs/2112.00749}{{\ttfamily 2112.00749}}].

\bibitem{Lange:2024pef}
S.C.~Lange et~al., \emph{{Galaxy mass modelling from multiwavelength JWST strong lens analysis: dark matter substructure, angular mass complexity, or both?}}, \href{https://doi.org/10.1093/mnras/staf491}{\emph{Mon. Not. Roy. Astron. Soc.} {\bfseries 539} (2025) 704} [\href{https://arxiv.org/abs/2410.12987}{{\ttfamily 2410.12987}}].

\bibitem{Vegetti:2023mgp}
S.~Vegetti et~al., \emph{{Strong Gravitational Lensing as a Probe of Dark Matter}}, \href{https://doi.org/10.1007/s11214-024-01087-w}{\emph{Space Sci. Rev.} {\bfseries 220} (2024) 58} [\href{https://arxiv.org/abs/2306.11781}{{\ttfamily 2306.11781}}].

\bibitem{Nightingale:2022bhh}
J.W.~Nightingale et~al., \emph{{Scanning for dark matter subhaloes in Hubble Space Telescope imaging of 54 strong lenses}}, \href{https://doi.org/10.1093/mnras/stad3694}{\emph{Mon. Not. Roy. Astron. Soc.} {\bfseries 527} (2023) 10480} [\href{https://arxiv.org/abs/2209.10566}{{\ttfamily 2209.10566}}].

\bibitem{Ephremidze:2025mqg}
N.~Ephremidze, C.~Chandrashekar, A.{\c{C}}.~{\c{S}}eng{\"u}l and C.~Dvorkin, \emph{{Dark Matter Substructure or Source Model Systematics? A Case Study of Cluster Lens Abell S1063}}, \href{https://doi.org/10.1093/mnras/staf1366}{\emph{Mon. Not. Roy. Astron. Soc.} {\bfseries 2610} (2025) 2638} [\href{https://arxiv.org/abs/2502.18571}{{\ttfamily 2502.18571}}].

\bibitem{Munoz:2019hjh}
J.B.~Mu{\~n}oz, C.~Dvorkin and F.-Y.~Cyr-Racine, \emph{{Probing the Small-Scale Matter Power Spectrum with Large-Scale 21-cm Data}}, \href{https://doi.org/10.1103/PhysRevD.101.063526}{\emph{Phys. Rev. D} {\bfseries 101} (2020) 063526} [\href{https://arxiv.org/abs/1911.11144}{{\ttfamily 1911.11144}}].

\end{thebibliography}\endgroup

\end{document}